\documentclass[%
 reprint,
superscriptaddress,
 amsmath,amssymb,
prx
]{revtex4-2}

\usepackage{graphicx}
\usepackage{dcolumn}

\usepackage[utf8]{inputenc} 
\usepackage[T1]{fontenc}    
\usepackage{hyperref}       
\usepackage{url}            
\usepackage{booktabs}       
\usepackage{amsfonts}       
\usepackage{nicefrac}       
\usepackage{microtype}      
\usepackage{xcolor}         
\usepackage{graphicx}
\usepackage{amsmath}
\usepackage[version=3]{mhchem}
\usepackage[normalem]{ulem}
\usepackage{braket}
\usepackage{tikz}
\usepackage{tikz}
\usetikzlibrary{arrows.meta}

\usepackage{listings}

\definecolor{codegreen}{rgb}{0,0.6,0}
\definecolor{codegray}{rgb}{0.5,0.5,0.5}
\definecolor{codepurple}{rgb}{0.58,0,0.82}
\definecolor{backcolour}{rgb}{0.95,0.95,0.92}

\lstdefinestyle{mystyle}{
    backgroundcolor=\color{backcolour},   
    commentstyle=\color{codegreen},
    keywordstyle=\color{magenta},
    numberstyle=\tiny\color{codegray},
    stringstyle=\color{codepurple},
    basicstyle=\ttfamily\footnotesize,
    breakatwhitespace=false,         
    breaklines=true,                 
    captionpos=b,                    
    keepspaces=true,                 
    numbers=left,                    
    numbersep=5pt,                  
    showspaces=false,                
    showstringspaces=false,
    showtabs=false,                  
    tabsize=2,
}

\DeclareMathOperator*{\argmin}{arg\,min}
\DeclareMathOperator{\G}{{\mathcal G}}

\DeclareMathOperator{\wG}{\widetilde{{\mathcal G}}}
\DeclareMathOperator{\F}{{\mathcal F}}
\DeclareMathOperator{\R}{\mathcal{R}}
\DeclareMathOperator{\vw}{\boldsymbol{w}}

\DeclareMathOperator{\vtheta}{\boldsymbol{\theta}}
\DeclareMathOperator{\vwp}{\boldsymbol{w}^{*}}

\usepackage[mathlines]{lineno}

\begin{document}

\preprint{APS/123-QED}

\title{Bottom-Up Design of Quantum Optical Experiments Using Discrete Generative Models}
\author{Isaac L. Huidobro-Meezs}
\email{huidobri@mcmaster.ca}
\affiliation{Department of Chemistry \& Chemical Biology, McMaster University, Hamilton, ON, Canada}
\author{Simón Paiva-Ortega}
\affiliation{Institut Courtois \& D\'epartement de physique, Universit\'e de Montr\'eal, Montréal, QC Canada}
\author{Rodrigo A. Vargas-Hernández}
\email{vargashr@mcmaster.ca}
\affiliation{Department of Chemistry \& Chemical Biology, McMaster University, Hamilton, ON, Canada}
\affiliation{School of Computational Science \& Engineering, McMaster University, Hamilton, ON  Canada}
\affiliation{Brockhouse Institute for Materials Research, McMaster University, Hamilton, ON, Canada}
\date{\today}

\begin{abstract}
Designing quantum optical experiments requires searching over discrete circuit topologies and continuous parameters, often leading to multiple possible realizations of the same target state. Graph-based methods commonly address this problem by optimizing a dense graph and pruning it toward a single circuit. We introduce \texttt{Grinch}, a bottom-up reward-driven generative framework that learns to sample optical graphs directly from fidelity-based rewards without relying on a pre-existing training dataset. We demonstrate the framework on multipartite entangled states, graph and cluster states, and nonlocal photonic-gate targets, obtaining multiple high-fidelity optical graphs. For nonlocal photonic gates, we obtain Toffoli constructions requiring two ancillas and alternative CNOT and Toffoli designs when connections between ancilla nodes are restricted. For state preparation, we identify asymptotic solutions for targets that cannot be generated exactly by graphs. Using the discovered topologies, we derive analytic families of optical experiments with explicit interference mechanisms and fidelity--rate relations, showing how generative design can support interpretable scientific discovery. These findings illustrate a reward-driven approach to quantum optical inverse design, exploring alternative circuit topologies while directly incorporating connectivity constraints into the search process.

\end{abstract}

\maketitle


\section{Introduction}

Photons are central to quantum technologies for computation, communication, sensing, metrology, and simulation. Realizing these applications often requires optical experiments that generate a prescribed quantum state. As the number of photons and accessible modes increases, the number of possible components, connections, and continuous parameters grows rapidly, motivating automated approaches to quantum-experiment design \cite{Melvin,TheseusDesign,PyTheus,DesignPhysExpAINature2026}. An important development for tackling this problem is the representation of quantum optical experiments as graphs \cite{QuantumExpAndGraphs,Mario_II,PossibleStates}, in which vertices represent optical paths, edges represent photon-pair sources, endpoint colors encode internal modes, and complex weights represent probability amplitudes. The graph's perfect matchings represent the multiphoton creation pathways whose coherent superposition determines the generated state. The graph therefore describes both the experimental setup and its output state, turning circuit design into a graph inverse problem.\\

Computer-aided design has adopted several strategies, including searches over sequences of optical components, genetic algorithms, active-learning agents that improve through feedback, and parametrized methods that optimize photonic circuits for state preparation and gate synthesis \cite{Melvin,KnottSearch,NicholsGenetic,RL-CircuitSearch,ArrazolaPhotonicML,InverseDesignPhotonicSystems}. More recently, graph-based approaches such as \texttt{Theseus} and \texttt{PyTheus}, which follow a \emph{top-down} approach, construct circuits for target states by progressively pruning edges from an initially fully connected graph \cite{TheseusDesign,PyTheus}. The use of Logic AI tools has also been considered \cite{Klaus}. These methods have produced high-dimensional entangled states and nonlocal photonic gates \cite{AI-entanglement,nonLocal-agents}. However, they also reveal two important features of the problem: i) useful circuits are usually sparse, and ii) several distinct circuits may generate the same target state. 
The latter has motivated recent work, including ours, using generative models, which are naturally suited to addressing this multiplicity of solutions. Previous studies have reconstructed quantum states, learned relationships between optical device sequences and entanglement, and generated new experiments using sequential graph models and variational autoencoders \cite{GenerativeQuantumStateReconstruction,LSTMQuantumOptics,SequentialGraphQuantumOptics,QOVAE}. However, these generative approaches require representative training data, which can be expensive to generate and can restrict the model to circuit families already present in the dataset. 

These limitations motivate a search strategy that learns directly from evaluations of candidate experiments and can accommodate different photonic tasks. Relevant targets include entangled states for quantum computing~\cite{Walther2005-nd}, spectroscopy~\cite{PhysRevLett.125.180502}, and quantum communication~\cite{PhysRevLett.84.4729}, as well as nonlocal photonic gates~\cite{nonLocal-agents}. For a given target, optical realizations can differ in the number of photon-pair sources and ancilla photons they require, and in the connections between optical paths. A search that incorporates these constraints and returns several high-fidelity designs would allow comparison of alternative implementations and adaptation to the resources of a particular experiment.

In this work, we introduce \texttt{Grinch}, a \textbf{bottom-up} framework based on generative flow networks (GFlowNets), which construct discrete objects and sample them with probability proportional to a chosen reward \cite{bengio2023gflownet,jain2023gflownets,zhang2023letFlows}.
Unlike existing top-down approaches that optimize and prune a single dense graph \cite{TheseusDesign,PyTheus,Klaus}, \texttt{Grinch} constructs circuit topologies edge by edge and learns directly from reward evaluations without requiring a pre-existing training dataset.
The discrete nature of GFlowNets has also been exploited in other applications in quantum sciences \cite{UsDiscreteFlowbased,PathIntegral}.
Within the framework for AI-driven experimental design proposed in Ref.~\cite{DesignPhysExpAINature2026}, \texttt{Grinch} addresses experimental-design problems involving both discrete structures and coupled discrete--continuous variables: the circuit topology is sampled discretely, while its continuous edge weights are optimized to maximize the reward.

\section{Methods}
\subsection{Quantum Optics Experiments as Graphs}
Photonic quantum experiments can be encoded in weighted colored graphs $\G(V, E)$, where the graph nodes ($V$) correspond to spatial photon paths, and the edges ($E$) between them encode the mode number of the photons through a specific color. The edge weight gives the amplitude associated with the photons. For example, the weight $w_{a,b}^{k_a,k_b}$ is encoded as the edge between photonic paths $a$ and $b$ with modes/colors $k_a$ and $k_b$ respectively. This can be translated into the operator $w_{a,b}^{k_a,k_b}a^\dagger_{k_a}b^\dagger_{k_b}+\mathrm{h.c.}$ with $a^\dagger_{k_a}$ the creation operator of a photon on path $a$ with mode $k_a$. The connection between the multi-graph and the resulting quantum state from the optical setup is given by a weight function \cite{PyTheus}
\begin{equation}
    \Phi(w) = \sum_m \frac{1}{m!}\left(\sum_{e\in E(\G)} w(e)x^\dagger(e)y^\dagger(e)+\text{h.c}\right)^m,
\end{equation}
where $E(\G)$ is the edge set, and $x^\dagger(e)$ and $y^\dagger(e)$ create photons in the paths connected by edge $e$, with internal modes specified by the corresponding endpoint colors. For an edge of the multi-graph $e=((a,b),(k_a,k_b))$, we have $x^\dagger(e)=a^\dagger_{k_a}$ and $y^\dagger(e)=b^\dagger_{k_b}$. Applying the weight function to the vacuum yields the resulting quantum state as a function of the edge weights. For a graph setup, the resulting quantum state corresponds to the superposition resulting from adding all perfect matchings of the graph, with each perfect matching generating a basis element corresponding to the modes incident to the paths \cite{Klaus}. The resulting graph representation can be translated into different photonic quantum-optics setups, including standard bulk optics, integrated photonics, and entanglement via path identity \cite{TheseusDesign}.
We refer the reader to Refs. \cite{QuantumExpAndGraphs, PossibleStates, Klaus, TheseusDesign,RL-CircuitSearch,nonLocal-agents,PyTheus} for more details regarding the mapping of quantum optical experiments to weighted colored graphs. 

\begin{figure}[h!]
    \centering
    \includegraphics[width=\linewidth]{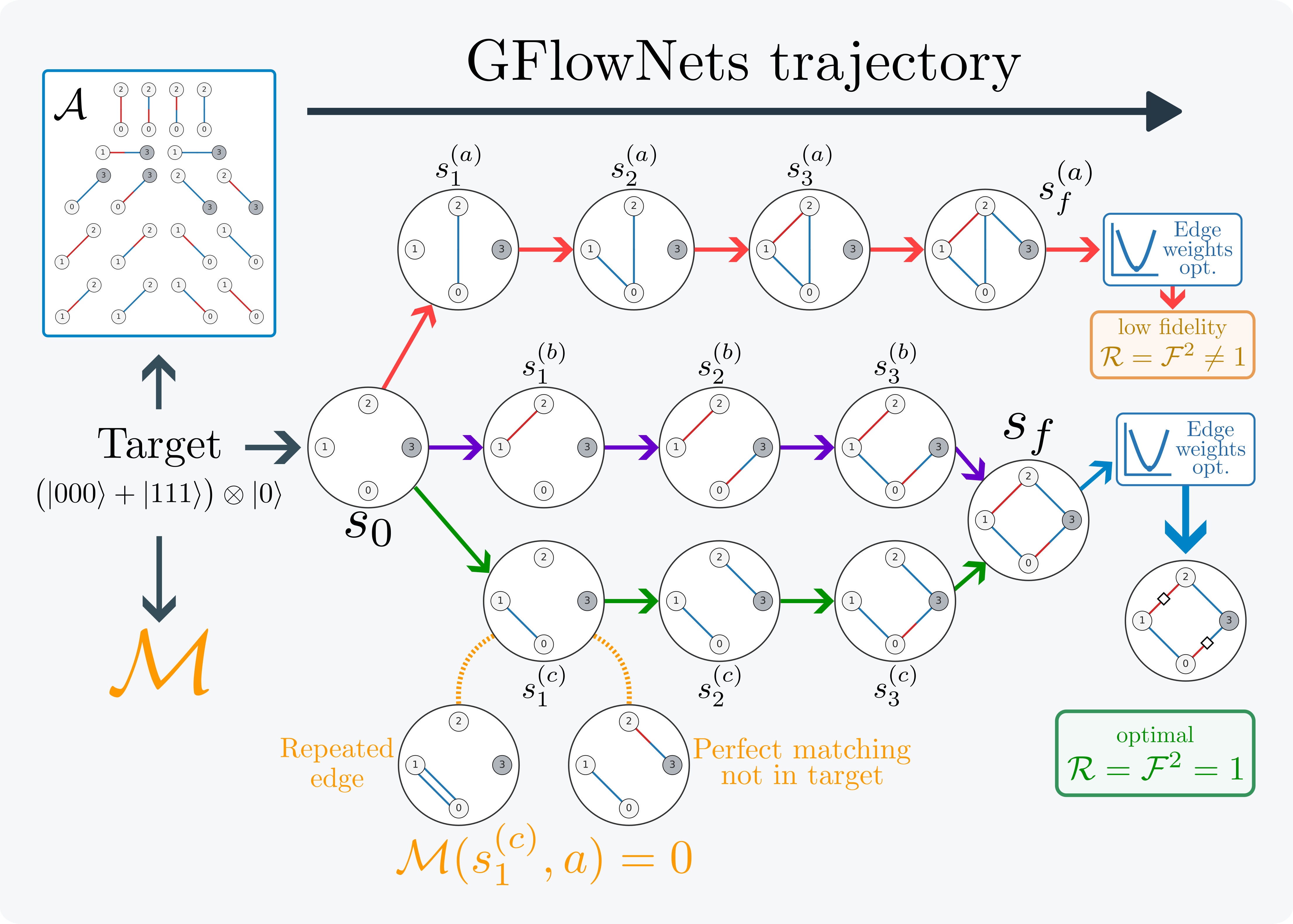}
    \caption[Bottom-up \texttt{Grinch} construction of GHZ(3,2).]{\textbf{Bottom-up} construction of
    $\ket{\mathrm{GHZ}(3,2)}\otimes\ket0$ with \texttt{Grinch}. The
    \tikz[baseline=-0.6ex]{\node[circle,
    draw={rgb,255:red,38;green,55;blue,70},
    fill={rgb,255:red,216;green,221;blue,225},line width=.5pt,
    minimum size=2.8mm,inner sep=0pt,font=\tiny\bfseries] {};} node is the ancilla.
    The mask filters out invalid photonic states and avoids perfect matchings that are not present in the target state.}
    \label{fig:grinch-ghz32}
\end{figure}

\subsection{Grinch}
A prominent class of graph-based inverse-design methods formulates the circuit topology and its continuous edge weights as a coupled discrete--continuous optimization problem \cite{TheseusDesign,PyTheus,Klaus},
\begin{equation}
(\wG,\vwp)\in
\argmin_{\mathcal{G},\vw}
\left[1-\mathcal{F}(\mathcal{G},\vw)\right],
\label{eqn:opt_fidelity}
\end{equation}
where \(\mathcal{F}(\mathcal{G},\vw)\) is the fidelity between the generated and target states, while \(\wG\) and \(\vwp\) denote an optimal topology (discrete) and its corresponding edge weights (continuous). Top-down methods begin with a dense graph and use optimized weights, often together with \(L_1\) regularization, to identify edges to remove \cite{TheseusDesign,PyTheus}. For \(n\) vertices and \(d\) colors, this dense graph contains $d^2n(n-1)/2$ weighted edges, excluding self-loops. In the common quantum state benchmarks, e.g., Greenberger-Horne-Zeilinger  (GHZ) and Schmidt-Rank-Vector (SRV) states, the target circuits retain only 5--10\% of these edges \cite{TheseusDesign,PyTheus,Klaus}. A top-down pruning procedure must therefore eliminate 90--95\% of the initial graph, whereas a bottom-up procedure starts from a disconnected graph and adds only a small number of edges. This motivates bottom-up construction as an alternative framework that may require fewer structural operations during the search.\\

In \texttt{Grinch}, we recast the discrete search as sampling circuit graphs from a reward-proportional distribution,
\begin{equation}
    \mathcal{G} \sim P_{\theta}(\mathcal{G})
    \qquad \text{where} \qquad
    P_{\theta}(\mathcal{G}) \propto \R(\mathcal{G}).  \label{eq:reward_sampling}
\end{equation}
$P_{\theta}(\mathcal{G})$ denotes the terminal distribution induced by a generative policy modeled through GFlowNets. Fig. \ref{fig:grinch-ghz32} provides a schematic of our proposed method. 
The model does not learn this distribution explicitly; instead, it learns construction policies whose terminal samples occur in proportion to the reward $\R(\mathcal{G})$. 
For quantum circuit design, the reward can be the fidelity or a combination of experimental metrics, while the continuous edge weights can still be optimized using standard methods. 
This sampling formulation is motivated by the existence of several inequivalent topologies with comparable fidelity, whereas Eq.~\ref{eqn:opt_fidelity} selects a single optimal circuit.\\

GFlowNets are designed to construct compositional objects through sequences of stochastic actions (policy). For a brief introduction to GFlowNets, see Appendix Sec. \ref{sec:app-gflownets}.
In \texttt{Grinch}, a state \(s\) is a partial circuit graph, the initial state \(s_0\) is an empty graph, and each transition \(s\rightarrow s'\) adds a colored edge from the set of actions $\mathcal{A}$ to produce the perfect matchings compatible with the target state.
A trajectory \(\tau=(s_0,\ldots,s_f=x)\) terminates at a completed circuit graph \(x\); see Fig.~\ref{fig:grinch-ghz32}. 
Each action adds one previously absent colored edge. A trajectory $\tau=(s_0,\ldots,s_{\ell_\tau}=s_f)$ terminates when it reaches the edge budget $T=\texttt{max\_edges}$ or when no admissible forward action remains, so its number of transitions satisfies $\ell_\tau\leq T$. For a target expansion with $K$ nonzero computational-basis terms on $N$ nodes, including ancillas, the quantity $NK/2$ serves as a heuristic reference for selecting the edge budget. 
The forward and backward policies, \(P_F\) and \(P_B\), are trained so that terminal graphs are sampled in proportion to their rewards, Eq.~\ref{eq:reward_sampling}. We use the trajectory-balance (TB) objective \cite{malkin2022trajectory},
\begin{equation}
\mathcal{L}_{\mathrm{TB}}(\tau)=\left(\log\frac{Z_\theta\prod_{t=1}^{\ell_\tau}P_F(s_t|s_{t-1};\vtheta)}{\mathcal{R}(s_f)\prod_{t=1}^{\ell_\tau}P_B(s_{t-1}|s_t;\vtheta)}\right)^2,
\label{eq:lossTB}
\end{equation}
where \(Z_\theta\) is a learnable estimate of the partition function and \(\R(x)\) is the reward of the terminal object. 
Minimizing Eq.~\ref{eq:lossTB} balances the probability of each forward construction trajectory with the reward-weighted probability of its reverse trajectory, allowing different trajectories to contribute to the same terminal graph.
For further details regarding GFlowNets, we refer the reader to Refs.~\cite{bengio2023gflownet,jain2023gflownets,malkin2022trajectory} and the Appendix Sec. \ref{sec:app-gflownets}.\\

In \texttt{Grinch}, we parameterize both policies, $P_F$ and $P_B$, and  \(\log Z_\theta\) with a UniMP graph transformer \cite{GraphTransformerPyTorch,GeneralizationTransformerOnGraphs}; three message-passing layers with four attention heads and hidden dimension 128 encode the node, edge, and connectivity information of each partial graph.  The quantity $\log Z_\theta$ is a separate trainable scalar optimized jointly with the policy parameters.
Global additive pooling produces a graph-level representation, which a two-layer multilayer perceptron maps to two sets of \(|\mathcal{A}|\) logits for the categorical forward and backward policies. 
Additional architectural and simulation details are provided in Appendix Sections \ref{sec:app-gflow-qoc} and \ref{sec:training_and_architecture}. \\

Throughout the trajectory, the action mask $\mathcal{M}(s,a)$ excludes duplicate edges, applies the configured endpoint-color and ancilla restrictions, and imposes perfect-matching completion bounds within the remaining edge budget. For quantum-gate targets, the candidate action set additionally excludes input--input edges. These restrictions reduce the search space, increasing the algorithm's efficiency by preferring valid and meaningful states. We employ a reward based on state fidelity; consequently, terminal candidates with fidelity $\mathcal{F}\geq0.95$ undergo greedy edge deletion to find simpler solutions within the edge budget. We accept a deletion only when it preserves a perfect matching for every required target component and maintains fidelity above the same threshold, with conditional reoptimization of the surviving weights; see Appendix Sec.~\ref{sec:masking_and_pruning} for details of the masking and pruning procedures.
Fig.~\ref{fig:grinch-ghz32} illustrates the resulting masked construction for $\ket{GHZ(3,2)}\otimes\ket{0}$.

\section{Results}
\subsection{Multipartite entangled states}
We evaluated \texttt{Grinch} on the automated design of quantum optical circuits for multipartite entangled photonic states, including high-dimensional Greenberger--Horne--Zeilinger (GHZ), Schmidt-rank-vector (SRV), and Dicke (D) states. Definitions of all target states are provided in Appendix Section \ref{sec:si_grinch_target_states}.
For each state, we train the GFlowNet by minimizing the TB objective, Eq.~\ref{eq:lossTB}, using Adam \cite{Adam} with a learning rate of \(10^{-3}\); see the Appendix Sec. \ref{sec:additional_results} for further details.
For each sampled terminal graph $s_f$ eligible for reward evaluation, we optimize its real edge weights using L-BFGS-B \cite{LBFGSB} with bounds $[-1,1]$ and assign the reward $\mathcal{R}(s_f)=\mathcal{F}(s_f,\vw_{\mathrm{opt}})^2$, where $\vw_{\mathrm{opt}}$ denotes the weights returned by the optimizer. Matching-completion calculations are cached for reuse during graph construction. We denote edges with negative amplitudes with a white diamond in the graph. 
We set the total number of samples to 5,000 for all multipartite entangled states and all subsequent calculations, as we observed that \texttt{Grinch} produced high-reward modes early in the sampling process, even when using an initial budget of 20,000 samples.
All the target-states notation in this work is defined in Appendix Sec.~\ref{sec:si_grinch_target_states}.

\begin{figure}[h!]
    \centering
    \includegraphics[width=0.8\linewidth]{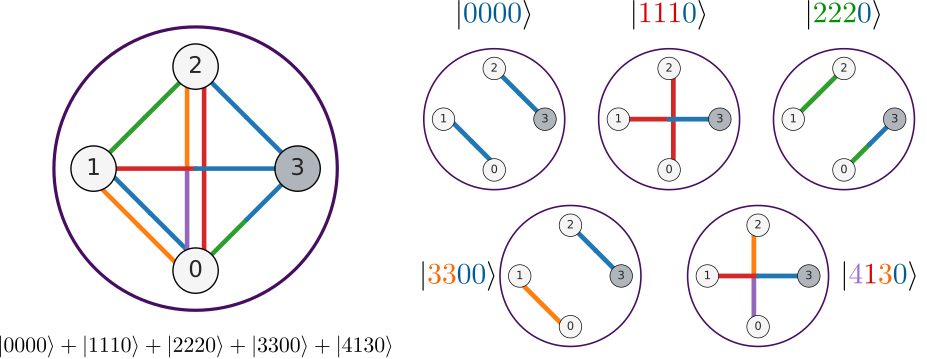}
    \caption[Best graph for the SRV(5,4,4) state.]{Best found graph for the $\ket{SRV(5,4,4)}\otimes\ket{0}$ state with its decomposition into perfect matchings. The  \tikz[baseline=-0.6ex]{\node[circle,
    draw={rgb,255:red,38;green,55;blue,70},
    fill={rgb,255:red,216;green,221;blue,225},line width=.5pt,
    minimum size=2.8mm,inner sep=0pt,font=\tiny\bfseries] {3$\,$};} corresponds to the ancilla photon detector.}
    \label{fig:GraphSRV544}
\end{figure}

\begin{table}[t]
\centering
\small
\setlength{\tabcolsep}{6pt}
\renewcommand{\arraystretch}{1.12}
\caption{Fidelity ($\mathcal{F}$) and optical-graph edge count ($|E|$) for multipartite entangled-state targets obtained with \texttt{Grinch}.
}
\label{tab:State-Fidelities}
\begin{tabular}{lcc}
\hline
\textbf{State} & $\mathcal{F}$ & $|E|$ \\
\hline
$\ket{GHZ(3,2)}\otimes\ket{0}$ & 1.00 & 4 \\
$\ket{GHZ(3,2)}\otimes\ket{0}$ & 1.00 & 4 \\
$\ket{GHZ(4,2)}$ & 1.00 & 4 \\
$\ket{GHZ(6,2)}$ & 1.00 & 6 \\
$\ket{GHZ(6,3)^*}$ & 1.00 & 9 \\
$\ket{GHZ(8,2)}$ & 1.00 & 8 \\
\hline
$\ket{SRV(5,4,4)}\otimes\ket{0}$ & 1.00 & 8 \\
$\ket{SRV(5,4,4)^*}\otimes\ket{0}$ & 0.80 & 18 \\
$\ket{SRV(5,4,4)^*}\otimes\ket{000}$ & 1.00 & 14 \\
$\ket{SRV(6,4,4)}\otimes\ket{0}$ & 1.00 & 9 \\
$\ket{SRV(6,4,4)^*}\otimes\ket{0}$ & 0.83 & 18 \\
$\ket{SRV(6,4,4)^*}\otimes\ket{000}$ & 1.00 & 15 \\
$\ket{SRV(6,5,4)}\otimes\ket{0}$ & 1.00 & 9 \\
$\ket{SRV(8,6,5)}\otimes\ket{0}$ & 1.00 & 12 \\
\hline
$\ket{D(3,(1,1,1))}\otimes\ket{0}$ & 1.00 & 9 \\
$\ket{D(4,(2,2))}$ & 1.00 & 8 \\
$\ket{W(4)}$ & 1.00 & 7 \\
\hline
\end{tabular}
\vspace{2pt}
\parbox{\linewidth}{\footnotesize
$^*$ Asymptotic solutions corresponding to states that cannot be generated exactly by optical graphs.
}
\end{table}

\begin{figure*}[]
    \centering

    \begin{minipage}[t]{0.45\textwidth}
        \centering
        \begin{tikzpicture}
            \node[anchor=south west, inner sep=0] (img) at (0,0)
                {\includegraphics[width=\linewidth]{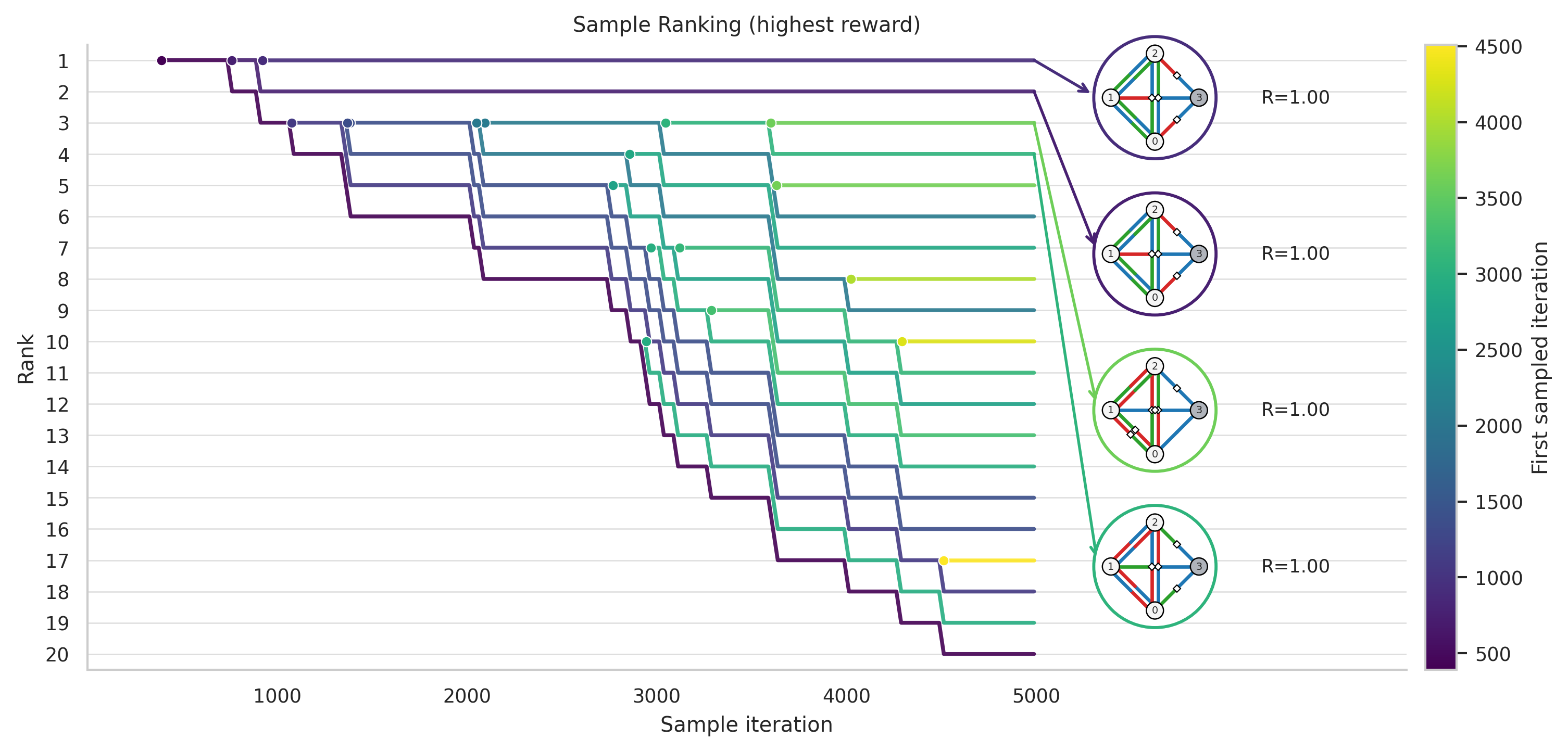}};
            \node[
                anchor=north west,
                fill=white,
                fill opacity=0.85,
                text opacity=1,
                inner sep=0.5pt,
                font=\scriptsize\bfseries
            ] at ([xshift=-10pt,yshift=-8pt]img.north west) {(a)};
        \end{tikzpicture}
    \end{minipage}
    \hfill
    \begin{minipage}[t]{0.45\textwidth}
        \centering
        \begin{tikzpicture}
            \node[anchor=south west, inner sep=0] (img) at (0,0)
                {\includegraphics[width=\linewidth]{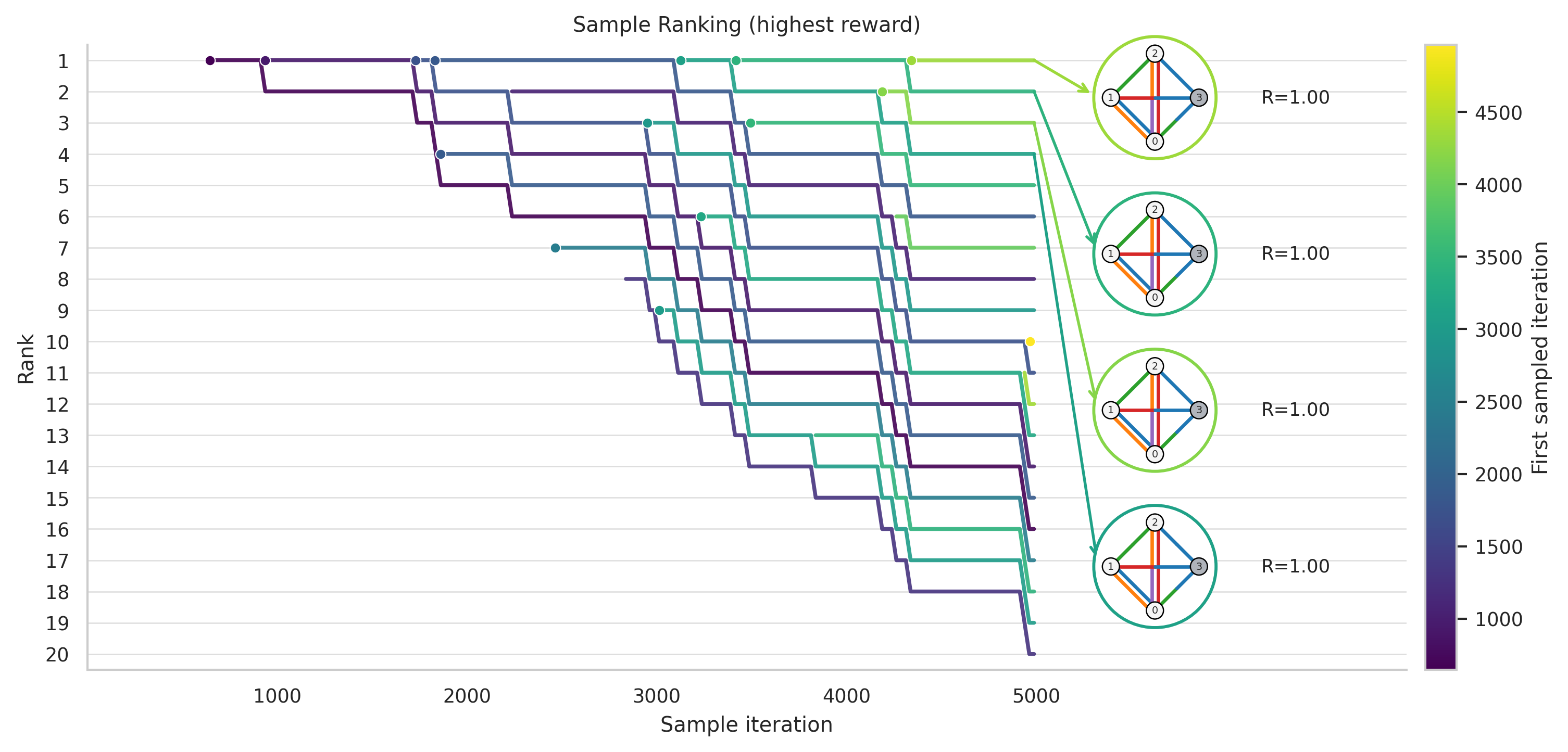}};
            \node[
                anchor=north west,
                fill=white,
                fill opacity=0.85,
                text opacity=1,
                inner sep=0.5pt,
                font=\scriptsize\bfseries
            ] at ([xshift=-10pt,yshift=-8pt]img.north west) {(b)};
        \end{tikzpicture}
    \end{minipage}

    \vspace{0.8em}

    \begin{minipage}[t]{0.45\textwidth}
        \centering
        \begin{tikzpicture}
            \node[anchor=south west, inner sep=0] (img) at (0,0)
                {\includegraphics[width=\linewidth]{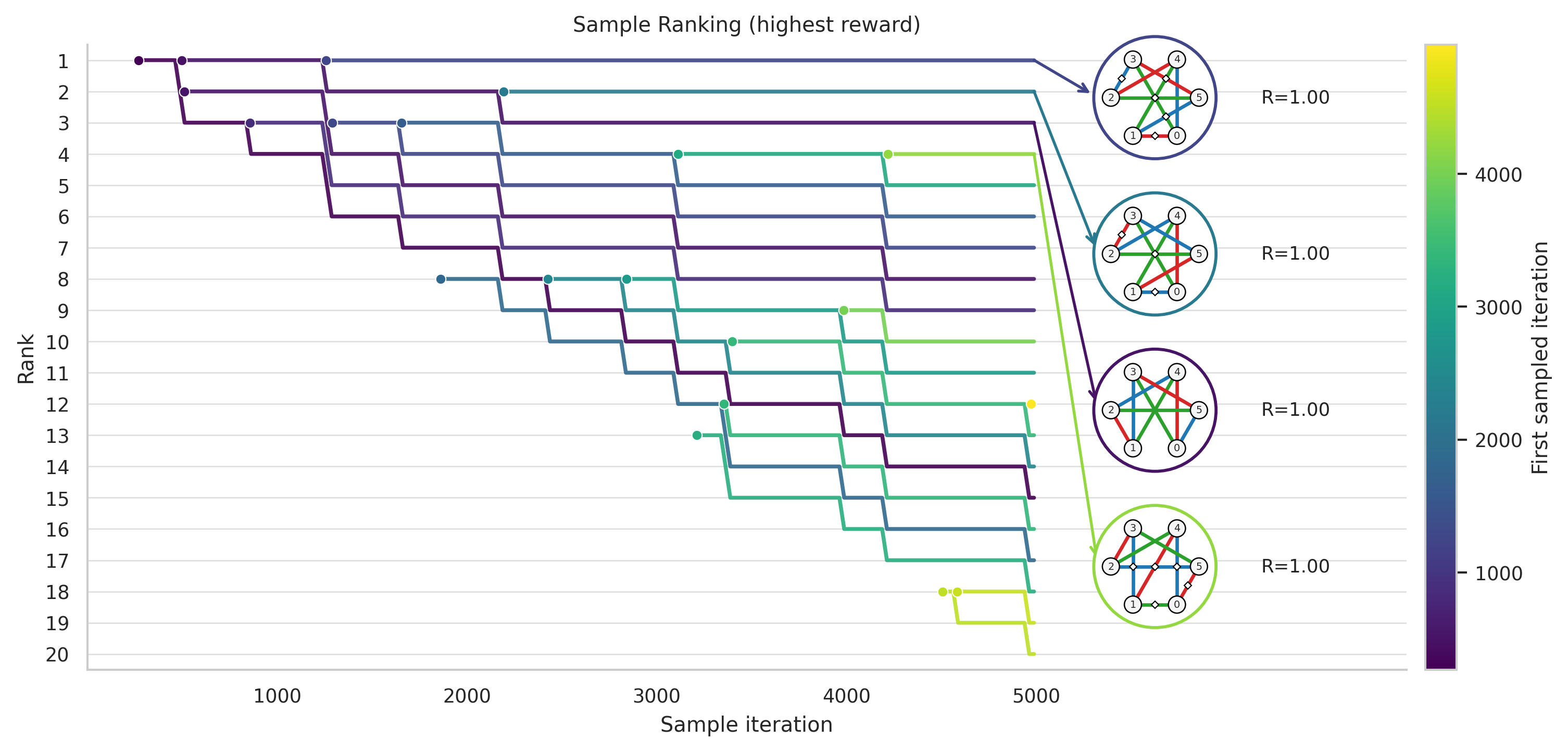}};
            \node[
                anchor=north west,
                fill=white,
                fill opacity=0.85,
                text opacity=1,
                inner sep=0.5pt,
                font=\scriptsize\bfseries
            ] at ([xshift=-10pt,yshift=-8pt]img.north west) {(c)};
        \end{tikzpicture}
    \end{minipage}
    \hfill
    \begin{minipage}[t]{0.45\textwidth}
        \centering
        \begin{tikzpicture}
            \node[anchor=south west, inner sep=0] (img) at (0,0)
                {\includegraphics[width=\linewidth]{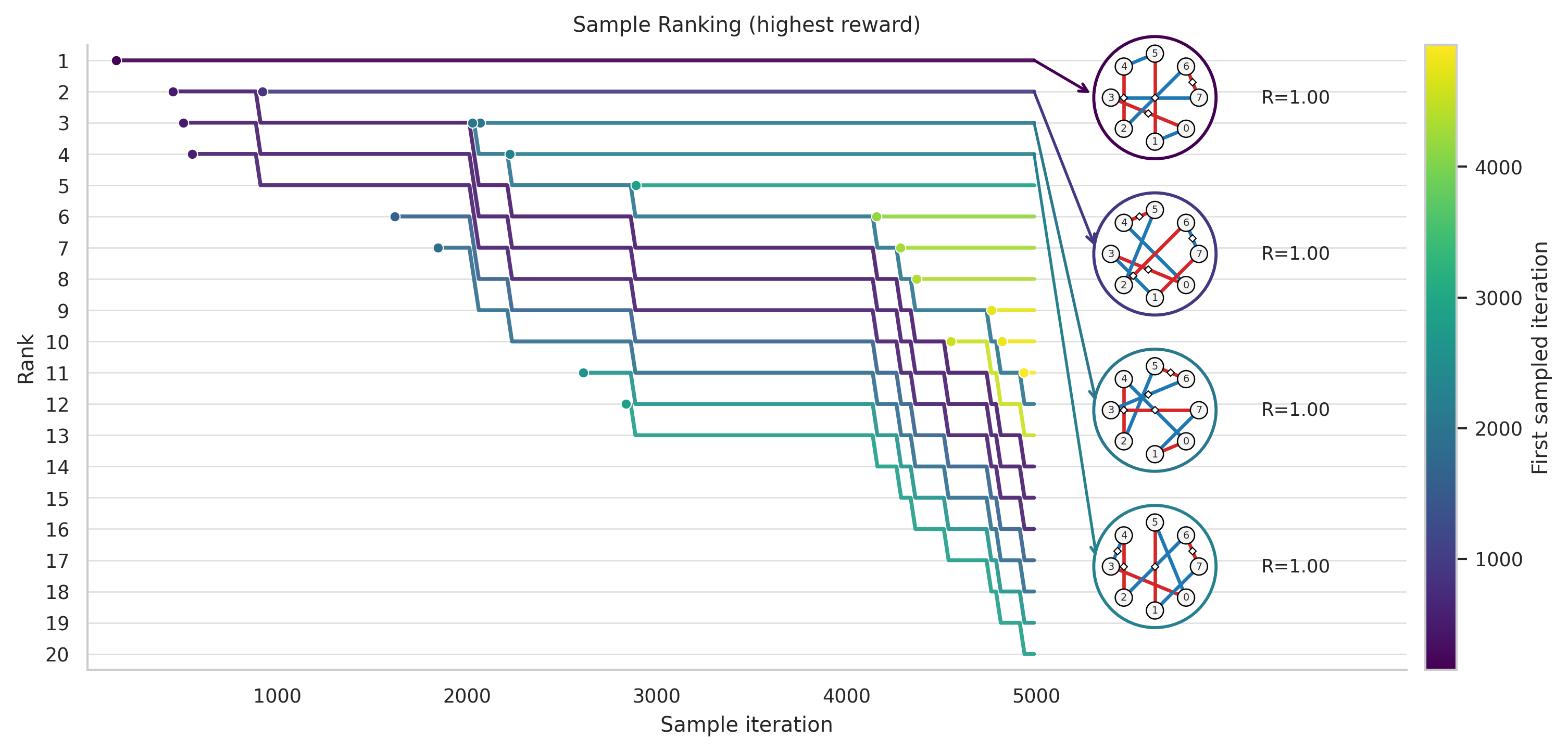}};
            \node[
                anchor=north west,
                fill=white,
                fill opacity=0.85,
                text opacity=1,
                inner sep=0.5pt,
                font=\scriptsize\bfseries
            ] at ([xshift=-10pt,yshift=-8pt]img.north west) {(d)};
        \end{tikzpicture}
    \end{minipage}

    \caption[Sample ranking over the number of iterations.]{Sample ranking over the number of iterations for (a) the $\ket{D(3,(1,1,1))}\otimes\ket{0}$, (b) $\ket{SRV(5,4,4)}\otimes\ket{0}$, (c) $\ket{GHZ(6,3)^*}$, and (d) $\ket{GHZ(8,2)}$ states colored by the first sampled iteration. The graphs of the top-4 solutions are shown as insets for each plot. The optimal solutions are found within the first quarter of the sample budget. 
    }
    \label{fig:Ranking_Ev}
\end{figure*}

We compare a smaller, strict edge budget with a larger, relaxed budget for the $\ket{SRV(5,4,4)}\otimes\ket{0}$ state, using $E_{min}=8$ as the baseline. The strict budget is $E_{min}+2$, and the relaxed budget is $E_{min}+4$. To exemplify the states found with \texttt{Grinch} and how they are decomposed into perfect matchings, we show in Fig. \ref{fig:GraphSRV544} the optimal sampled circuit for the $\ket{SRV(5,4,4)}\otimes\ket{0}$ state. From this strict-vs-relaxed comparison, we found that allowing \texttt{Grinch} to sample circuits with more edges reduces the overall number of samples required to discover high-reward solutions compared to the strict campaign. 
For example, in Fig.~\ref{fig:rewards_progress_10vs12} in Appendix Sec. \ref{sec:additional_results}, we tested our approach on the $\ket{SRV(5,4,4)}\otimes\ket{0}$ state without the logical masking functions to isolate the effect of the trajectory length. We observe that the relaxed campaign finds a unit-fidelity solution in fewer than 3,000 samples, whereas the strict campaign requires more than 12,000 samples to obtain a similar solution. Furthermore, with the masking functions employed in \texttt{Grinch}, the number of samples necessary to reach unit fidelity decreases dramatically to less than 1,000 samples for the same state, as shown in Fig. \ref{fig:Ranking_Ev}b.\\

Table \ref{tab:State-Fidelities} reports the resulting fidelities and edge counts for the solutions found with \texttt{Grinch}, which correctly constructs high-dimensional GHZ states with $\mathcal{F}\approx 1$. The $\ket{GHZ(6,3)^*}$ state cannot be prepared perfectly with linear optics and probabilistic photon-pair sources \cite{QuantumExpAndGraphs}. 
However, the resulting circuit can be implemented experimentally, as its fidelity scales as $1-\mathcal{O}(\vw^4)$, consistent with previously reported arrays \cite{TheseusDesign}. The Dicke states $\ket{W(4)}$ and $\ket{D(4,(2,2))}$ are recovered with seven and eight edges, respectively, extending the state-preparation results to superpositions with one and two excitations. 
For states that require an ancilla photon, which are key to getting rid of Maverick terms and to solve graphs that have an odd number of states, the set of actions, $\mathcal{A}$, is augmented with bicolored weighted edges between the ancilla node $v_A$ with mode number $k_A=0$ and the regular nodes in the graph, $v_i$, with mode number $k_i$. This allows us to generate, within the same framework, any state $|\psi\rangle_{Target}\otimes|0\rangle$. The first two examples of this are the $\ket{GHZ(3,2)}\otimes\ket{0}$ and the $\ket{D(3,(1,1,1))}\otimes\ket{0}$ Dicke states, which require an ancilla node as they entangle an odd number of photonic paths. \texttt{Grinch} allows for facile inclusion of any number of ancilla nodes and mode numbers, handling general states $|\psi\rangle_{Target}\otimes|\psi\rangle_{Ancilla}$, making it comparable to PyTheus \cite{PyTheus} and enabling exploration of more complicated optical arrays. \texttt{Grinch} also allows adding restrictions on the set of actions for hardware-restricted arrays, where two nodes are prevented from forming an edge. Table~\ref{tab:State-Fidelities-vs} compares \texttt{Grinch} with four reference frameworks for the seven shared target states reported in Ref.~\cite{Klaus}. We observe that \texttt{Grinch} can match the solutions found by such approaches while maintaining a consistent number of edges across the solutions. \\

\begin{figure}[t!]
    \centering
    \includegraphics[width=\linewidth]{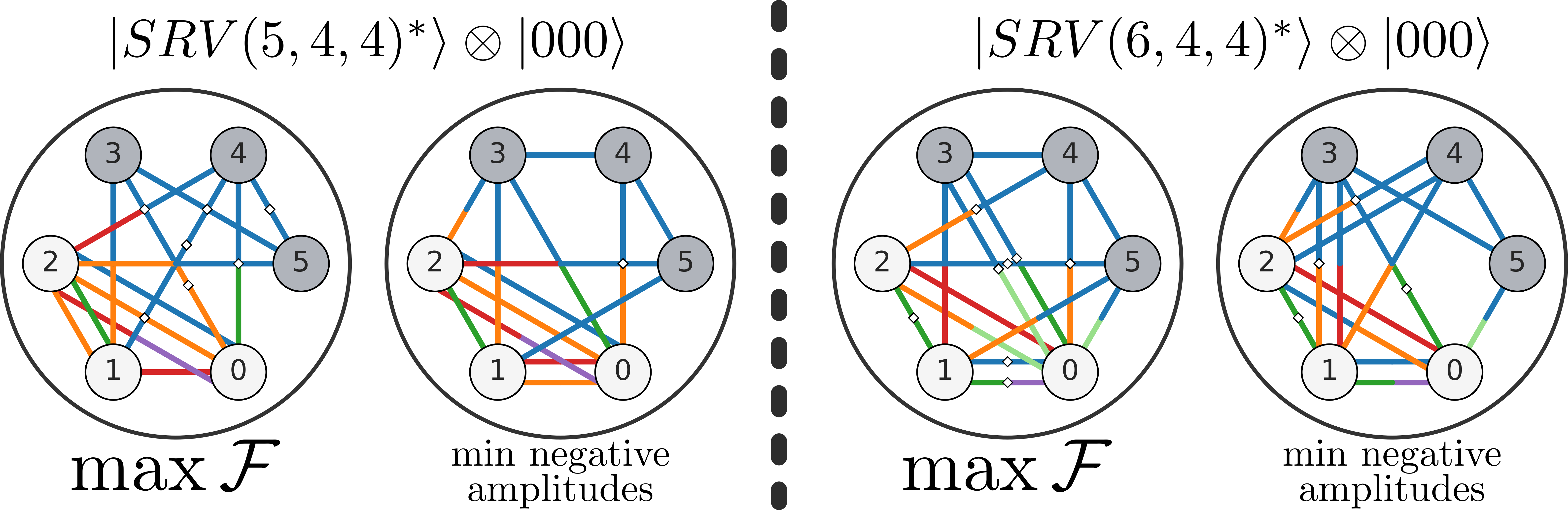}
    \caption{Graphs found for the $\ket{SRV(5,4,4)^*}\otimes\ket{000}$ and $\ket{SRV(6,4,4)^*}\otimes\ket{000}$ SRV states. The greyed-out node corresponds to the ancilla photon detector. Edges with negative amplitudes are signaled with a white diamond. Intruder states are reduced in magnitude through the weight optimization.}
    \label{fig:SRVs3a}
\end{figure}

Within the SRV benchmarks, we encounter the first targets that cannot be produced with unit fidelity, namely the states $\ket{SRV(5,4,4)^*}\otimes\ket{0}$ and $\ket{SRV(6,4,4)^*}\otimes\ket{0}$. The maximum fidelity found with our framework for these states is consistent with previously reported values, as these states cannot be exactly generated by graphs \cite{Klaus}. To showcase the flexibility of \texttt{Grinch}, we explore this state preparation by considering an array with two extra ancilla nodes, $\ket{SRV(5,4,4)^*}\otimes\ket{000}$ and $\ket{SRV(6,4,4)^*}\otimes\ket{000}$. Fig. \ref{fig:SRVs3a} shows the graphs found for these states, labeling the state with the highest asymptotic fidelity and the one with the fewest negative amplitudes. In both cases, we can find solutions that achieve asymptotically unit fidelity, with the intruder states reduced via weight optimizations, as in the $\ket{GHZ(6,3)^*}$ case. For $\ket{SRV(5,4,4)^*}\otimes\ket{000}$, the highest fidelity state approaches unity within a $10^{-6}$ tolerance, while the best state with the lowest number of negative amplitudes reaches a tolerance of $10^{-6}$. For $\ket{SRV(6,4,4)^*}\otimes\ket{000}$, the highest fidelity and lowest negative amplitude solutions achieve $10^{-6}$ and $10^{-5}$ tolerances respectively. This is an example of the added benefit of the candidate diversity produced through GFlowNets with \texttt{Grinch}. As reported in Table \ref{tab:State-Fidelities}, \texttt{Grinch} can find the other SRV states considered in the benchmark with unit fidelity.\\

The topologies identified by \texttt{Grinch} also enable explicit analytic constructions of optical experiments. For the 14- and 15-edge graphs labeled $\max\mathcal{F}$ in Fig.~\ref{fig:SRVs3a}, the one-parameter edge-weight assignments derived in Appendix Sec.~\ref{sec:srv-constructions} retain every colored edge for $0<t\leq1$. Destructive interference cancels two intruder components in each construction, leaving three and two residual intruders, respectively, each with amplitude $-t^2/2$. The resulting unnormalized postselected states are
\begin{equation}
\begin{aligned}
\ket{\widetilde\psi_{544^*}(t)}={}&\sqrt{5}\,t\ket{SRV(5,4,4)^*}\otimes\ket{000}\\
&-\frac{t^2}{2}\bigl(\ket{030000}+\ket{322000}+\ket{431000}\bigr),\\[3pt]
\ket{\widetilde\psi_{644^*}(t)}={}&\sqrt{6}\,t\ket{SRV(6,4,4)^*}\otimes\ket{000}\\
&-\frac{t^2}{2}\bigl(\ket{131000}+\ket{233000}\bigr).
\end{aligned}
\label{eq:srv-raw}
\end{equation}
Here the SRV target kets are normalized, so each target basis component has amplitude $t$. The corresponding conditional fidelities are $\mathcal{F}_{544^*}(t)=20/(20+3t^2)$ and $\mathcal{F}_{644^*}(t)=12/(12+t^2)$, both approaching unity as $t\to0^+$. This limit entails a tradeoff between fidelity and preparation rate. Within the weak-pair-generation model, at fixed common source scale, repetition rate, and mode-independent transmission and detection efficiencies, the detected sixfold coincidence rates along these families satisfy $R_{\alpha}(t)\propto[1-\mathcal{F}_{\alpha}(t)]/\mathcal{F}_{\alpha}(t)^2$ with $\alpha\in\{544^*,644^*\}$. The rates therefore vanish linearly with infidelity as the conditional fidelities approach unity. These constructions connect the topologies discovered by \texttt{Grinch} to analytic optical designs with explicit interference mechanisms and fidelity--rate relations.

We next examine how efficiently \texttt{Grinch} discovers high-fidelity graphs and the diversity of the sampled solutions. Fig. \ref{fig:Ranking_Ev} shows the ranking evolution of the top-20 unique trajectory graphs found by \texttt{Grinch} during the sampling process for the states $\ket{D(3,(1,1,1))}\otimes\ket{0}$, $\ket{SRV(5,4,4)}\otimes\ket{0}$, $\ket{GHZ(6,3)^*}$, and $\ket{GHZ(8,2)}$. We found that \texttt{Grinch} needed fewer than 1,000 samples to produce a unit-fidelity graph in all cases. Fig. \ref{fig:Ranking_Ev}-(a, c, d) illustrates the added benefit of sampling diversity, as \texttt{Grinch} can obtain several unique competitive solutions. In other cases, like $\ket{SRV(5,4,4)}\otimes\ket{0}$, different trajectories with distinct final states can lead to the same pruned state as shown in the figure. Given the bottom-up approach, the sampling process is efficient, with potential extensions related to parallelized sampling, and can be stopped once a desired fidelity is reached, allowing for efficient exploration of the solution space; see Fig.~\ref{fig:timing} in the Appendix Section \ref{sec:sampling-rates} for the sampling rates as a function of problem dimension.

\begin{figure}[ht!]
    \centering

    \begin{minipage}[t]{0.37\textwidth}
        \centering
        \begin{tikzpicture}
            \node[anchor=south west, inner sep=0] (img) at (0,0)
                {\includegraphics[width=\linewidth]{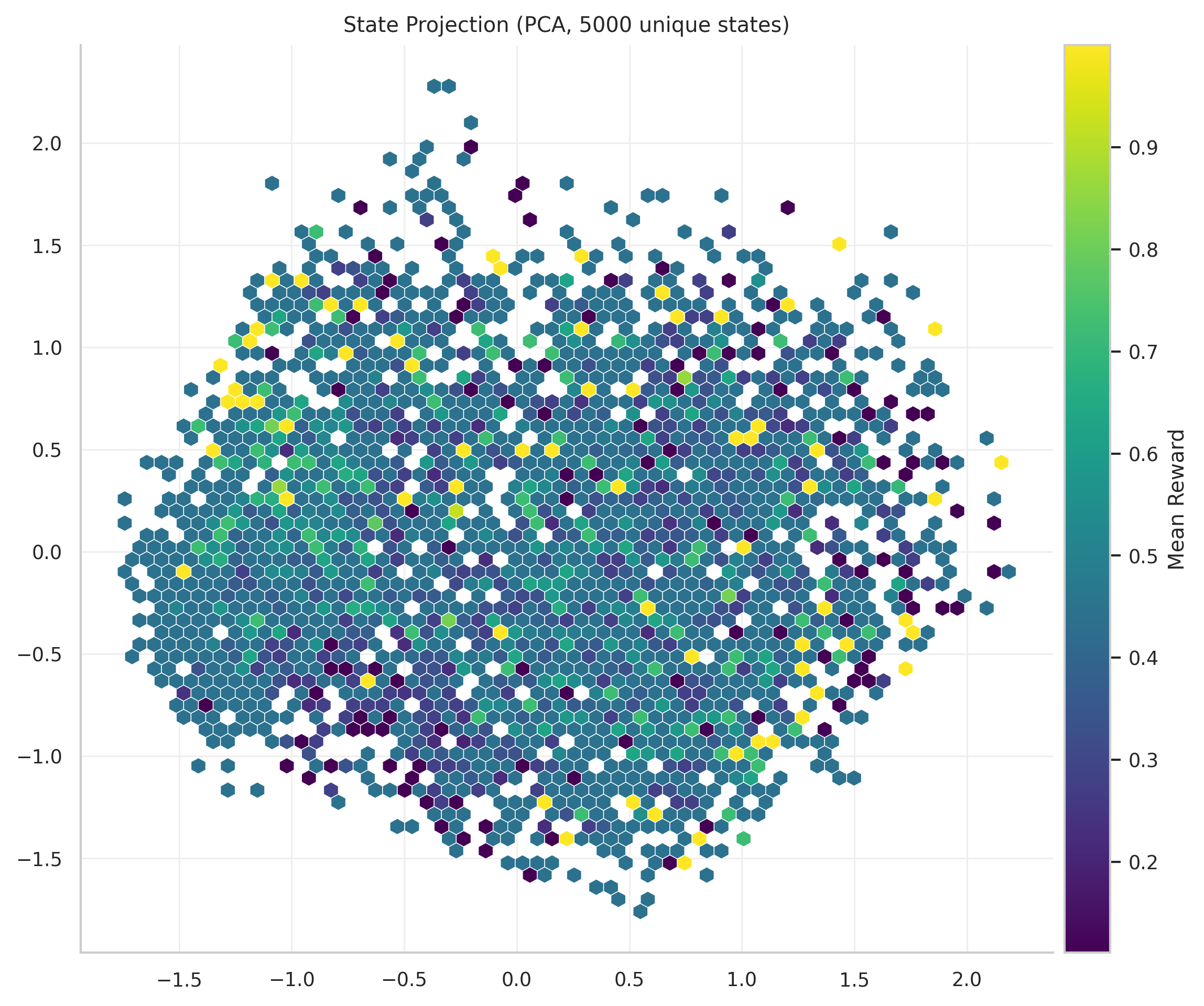}};
            \node[
                anchor=north west,
                fill=white,
                fill opacity=0.85,
                text opacity=1,
                inner sep=0.5pt,
                font=\scriptsize\bfseries
            ] at ([xshift=-10pt,yshift=-8pt]img.north west) {(a)};
        \end{tikzpicture}
    \end{minipage}

    \begin{minipage}[t]{0.37\textwidth}
        \centering
        \begin{tikzpicture}
            \node[anchor=south west, inner sep=0] (img) at (0,0)
                {\includegraphics[width=\linewidth]{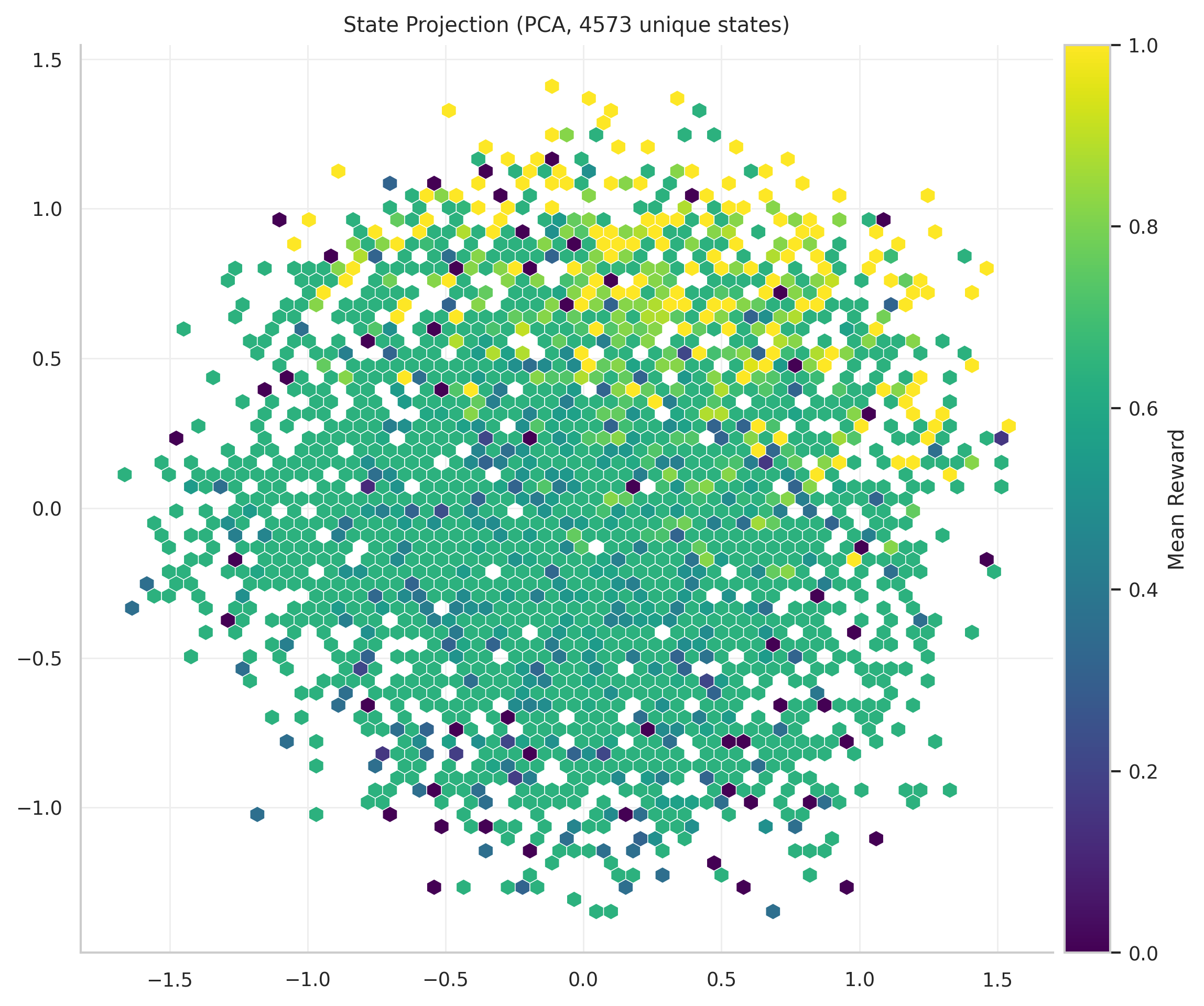}};
            \node[
                anchor=north west,
                fill=white,
                fill opacity=0.85,
                text opacity=1,
                inner sep=0.5pt,
                font=\scriptsize\bfseries
            ] at ([xshift=-10pt,yshift=-8pt]img.north west) {(b)};
        \end{tikzpicture}
    \end{minipage}

    \caption[Principal component analysis of the sampled graphs.]{Principal component analysis of the sampled graphs for (a) the $\ket{GHZ(6,3)^*}$ and (b) the $\ket{SRV(5,4,4)}\otimes\ket{0}$ states.}
    \label{fig:PCA}
\end{figure}

To gain insight into the potential benefits of the diverse sampling of GFlowNets, we performed a principal component analysis (PCA) and plotted the distribution of states over the first two components in Fig. \ref{fig:PCA}, colored by the average reward on a hexagonal grid. We selected the $\ket{GHZ(6,3)^*}$ and $\ket{SRV(5,4,4)}\otimes\ket{0}$ states as examples, both with the same edge budget of fourteen, as their PCA projections exemplify two distinct regimes. For $\ket{GHZ(6,3)^*}$, several possible asymptotic optical arrays with unit fidelity are possible, leading to different localized regions on the PCA two-dimensional projection with high rewards. However, a budget closer to the number of edges in the optimal solution yields more localized reward regions as the sampling space is reduced. On the other hand, for a larger difference from the optimal configuration, such as $\ket{SRV(5,4,4)}\otimes\ket{0}$, the high reward region is spread across the two-dimensional projection. This behavior results from the larger number of possible sampled states that, after pruning, yield unit-fidelity arrays.

These examples illustrate how diverse sampling can be beneficial. When multiple arrays are possible, each with potential hardware differences, GFlowNets can effectively explore these regions and identify multiple high-reward modes. When the solution is concentrated in a specific section of the solution space, the sampling process allows to identify that mode; however, for top-down approaches, such problems might be more difficult, as this means that if a pruning step in the topological optimization removed an edge which is necessary, the end result would not have the best possible fidelity, explaining the high variance in these approaches \cite{Klaus}.

\subsection{Graph states}
\label{sec:results_graph_states}

Photonic graph and cluster states are resources for measurement-based quantum computation, in which a computation is implemented through measurements on a previously prepared entangled state rather than a sequence of gates acting on initially uncorrelated qubits \cite{Raussendorf2003Cluster}. Four-photon cluster states have been used experimentally to demonstrate this approach \cite{Walther2005OneWay}. Small graph states also play a role in quantum error correction: fusion-based photonic architectures combine entangled resource states via joint measurements, whose outcomes serve as parity checks for fault-tolerant computation. These architectures include constructions based on four-qubit star resources \cite{Bartolucci2023Fusion}. These applications motivate the search for compact photonic state-generation graphs.

Table~\ref{tab:graph_states} shows the resulting edge counts found with \texttt{Grinch} for six graph states, all with unit fidelity. For the explicit target state definitions, we refer the reader to Appendix Sec. \ref{sec:si_grinch_target_states}. 
We report the smallest recorded graph, when an accepted output of the pruning procedure is available, using the reoptimized fidelity to break ties in edge count. Otherwise, we report the highest-fidelity unpruned solution. Reported fidelities are obtained by reoptimizing the weights of the stored topologies during postprocessing.

\begin{table}[t]
\centering
\small
\setlength{\tabcolsep}{6pt}
\renewcommand{\arraystretch}{1.12}
\caption{
Fidelity ($\mathcal{F}=1$) and optical-graph edge count ($|E|$) for the smallest pruned solutions found by \texttt{Grinch}. 
}
\label{tab:graph_states}
\begin{tabular}{lcc}
\hline
\textbf{Target State} & $\mathcal{F}$ & $|E|$ \\
\hline
$\ket{G(2)}$ & 1.00 & 4 \\
$\ket{G_{\mathrm{lin}}(3)}\otimes\ket{0}$ & 1.00 & 8 \\
$\ket{G_{\mathrm{lin}}(4)}$ & 1.00 & 15 \\
$\ket{G_{\mathrm{star}}(4)}$ & 1.00 & 12 \\
$\ket{C_{\mathrm{1D}}(4)}$ & 1.00 & 7 \\
$\ket{C_{\mathrm{2D}}(4)}$ & 1.00 & 14 \\
\hline
\end{tabular}
\end{table}

\begin{figure*}[ht!]
    \centering
    \includegraphics[width=0.98\linewidth]{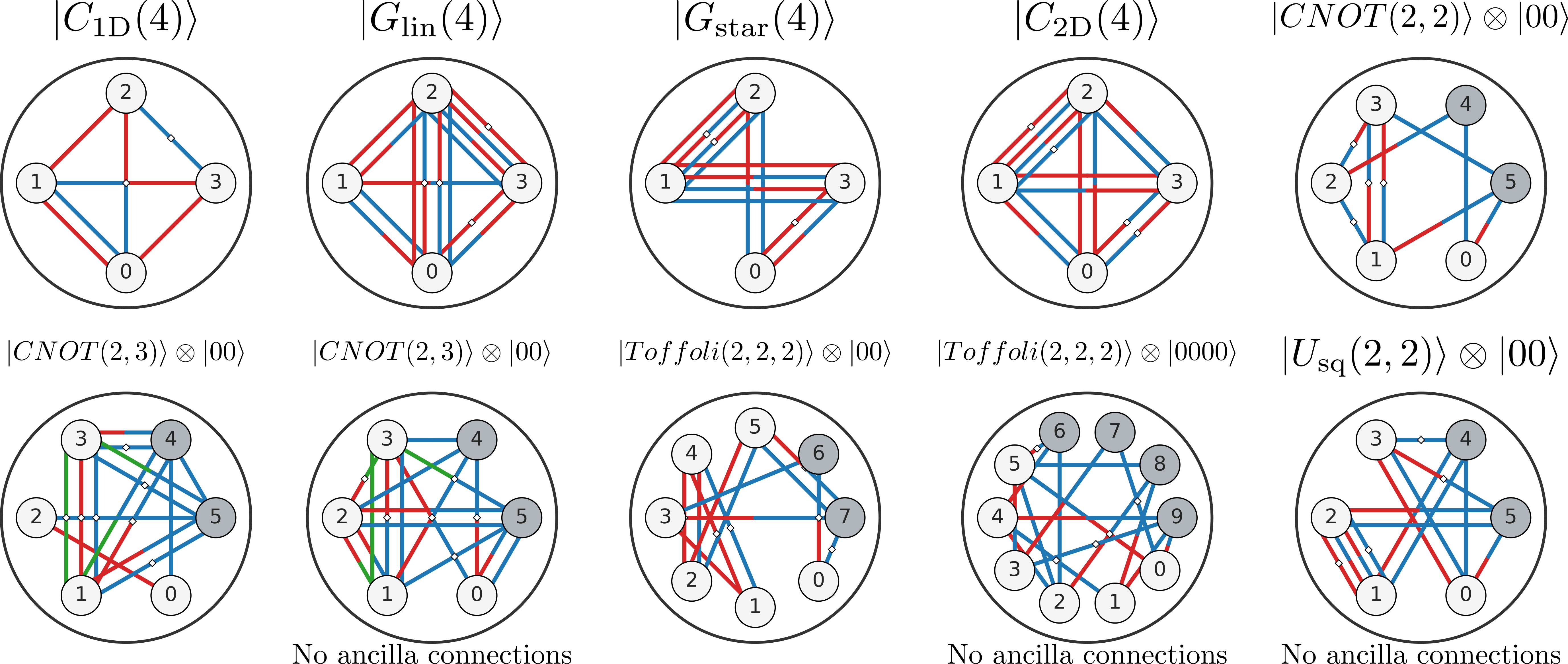}
    \caption[Best graphs for different states. ]{Best found graphs for different graph states and nonlocal photonic gate designs found with \texttt{Grinch}. Edges with negative amplitudes are signaled with a white diamond. The  \tikz[baseline=-0.6ex]{\node[circle,
    draw={rgb,255:red,38;green,55;blue,70},
    fill={rgb,255:red,216;green,221;blue,225},line width=.5pt,
    minimum size=2.8mm,inner sep=0pt,font=\tiny\bfseries] {$\,$};} corresponds to the ancilla photon detector.}
    \label{fig:GraphStatesGates}
\end{figure*}

The two-vertex graph state $\ket{G(2)}$ is recovered with four optical edges. Among the four-qubit graph and cluster targets, the smallest recorded solutions contain seven edges for $\ket{C_{\mathrm{1D}}(4)}$, twelve for $\ket{G_{\mathrm{star}}(4)}$, fourteen for $\ket{C_{\mathrm{2D}}(4)}$, and fifteen for $\ket{G_{\mathrm{lin}}(4)}$. 
The distinction between target-state entanglement and its optical representation is important when interpreting these counts. For the target conventions used here, the four-term cluster state $\ket{C_{\mathrm{1D}}(4)}$ and the sixteen-term linear graph state $\ket{G_{\mathrm{lin}}(4)}$ are related by Hadamard transformations on the two end qubits \cite{LinearClusterEquiv}. Their seven- and fifteen-edge solutions therefore illustrate that the optical graphs found by the search can depend on the chosen local-basis representation, even for locally equivalent targets. This comparison does not include the optical elements needed to implement those local basis changes and does not establish globally minimal resource requirements. We show the found graphs in Fig. \ref{fig:GraphStatesGates} for the relevant graph states and nonlocal photonic gates discussed in the following section.

\subsection{Nonlocal photonic gates}
\label{sec:results_photonic_gates}

Entangling photonic gates are building blocks for optical quantum computation. Linear-optical schemes can realize effective interactions through interference, ancillary photons, and conditional measurements, avoiding the need for strong direct photon--photon coupling \cite{Knill2001LinearOptics}. Extending such operations to spatially separated systems is relevant to distributed quantum processing, in which nonlocal gates connect distinct nodes in quantum networks \cite{Liu2024Nonlocal}. The synthesis of CNOT, CZ, and multiply controlled transformations therefore extends the design task from preparing entangled resources to reproducing coherent logical operations.

For non-local photonic gates, we represent the desired action of the gate through the following encoding of the desired input--output relation as target state
\begin{equation}
    |\text{in}\rangle\otimes|\text{out}\rangle = |\text{in, out}\rangle
\end{equation}
with the condition that the input qubits can not share an edge between them. For example, given a CNOT gate with transformation
\begin{align*}
    \ket{00}\to\ket{00}\\
    \ket{01}\to\ket{01}\\
    \ket{10}\to\ket{11}\\
    \ket{11}\to\ket{10}
\end{align*}
the target state is 
\begin{equation}
    \ket{00,00}+\ket{01,01}+\ket{10,11}+\ket{11,10}
\end{equation}
and the set of actions is reduced because incoming photons cannot share edges \cite{PyTheus}. \texttt{Grinch} takes the target state and, through a flag indicating that the photonic gate generation task is active, reduces the set of actions accordingly.

\begin{table}[t]
\centering
\small
\setlength{\tabcolsep}{6pt}
\renewcommand{\arraystretch}{1.12}
\caption{
Fidelity ($\mathcal{F}$) and optical-graph edge count ($|E|$) for nonlocal quantum-gate targets obtained with \texttt{Grinch}. Unmarked entries correspond to the smallest recorded pruned solutions and their associated fidelities. $\dagger$ denotes solutions with restricted edges between ancilla nodes. 
}
\label{tab:quantum_gates}
\begin{tabular}{lcc}
\hline
\textbf{Target State} & $\mathcal{F}$ & $|E|$ \\
\hline
$\ket{CZ(2,2)}$ & 1.00 & 6 \\
$\ket{CZ(2,2)}\otimes\ket{00}$ & 1.00 & 8 \\
$\ket{CNOT(2,2)}\otimes\ket{00}$ & 1.00 & 9 \\
$\ket{CNOT(2,3)}\otimes\ket{00}$ & 1.00 & 15/17$^\dagger$ \\
$\ket{U_{\mathrm{sq}}(2,2)}\otimes\ket{00}$ & 1.00 & 13/14$^\dagger$ \\
$\ket{Toffoli(2,2,2)}\otimes\ket{00}$ & 1.00 & 13 \\
$\ket{Toffoli(2,2,2)}\otimes\ket{0000}$ & 1.00 & 15/$17^\dagger$ \\
\hline
\end{tabular}
\end{table}

As shown in Table \ref{tab:quantum_gates}, the targets $\ket{CNOT(2,2)}\otimes\ket{00}$ and $\ket{CZ(2,2)}\otimes\ket{00}$ reach fidelity $1.00$, with the smallest recorded pruned graphs containing nine and eight edges, respectively. The ancilla-free target $\ket{CZ(2,2)}$ also reaches this fidelity, with a six-edge graph. The mixed-dimensional target $\ket{CNOT(2,3)}\otimes\ket{00}$, which encodes a gate that conditionally increments a qutrit target modulo three under the control of a qubit, reaches fidelity $1.00$ with a fifteen-edge pruned graph. To test \texttt{Grinch} under hardware constraints, we implemented a test in which the ancilla nodes were not allowed to communicate directly. This led to an alternative solution with seventeen edges; to see the difference between the arrays, we show both solutions in Fig. \ref{fig:GraphStatesGates}. 

We searched for constructions of nonlocal Toffoli gates requiring fewer ancillas. The two-ancilla Toffoli target is accurately reproduced under the tested settings. The target $\ket{Toffoli(2,2,2)}\otimes\ket{00}$ reaches fidelity $1.00$ with a thirteen-edge pruned graph. The four-ancilla counterpart can achieve unit fidelity using constructions equivalent to previously reported arrays \cite{nonLocal-agents}. In close analogy to the hardware restrictions tested for $\ket{CNOT(2,3)}\otimes\ket{00}$, we tested a four-ancilla Toffoli gate $\ket{Toffoli(2,2,2)}\otimes\ket{0000}$, finding a solution with seventeen edges under this restriction, while the unrestricted version requires fifteen. 

We also consider the two-qubit unitary $U_{\mathrm{sq}}(2,2)=CZ(H\otimes H)CZ$, whose encoded input--output target has the square-cluster expansion $\ket{C_{\mathrm{2D}}(4)}$ when the first two nodes are assigned to the input register and the last two to the output register. The gate maps $\ket{00}$ to $\ket{G(2)}$. For the target $\ket{U_{\mathrm{sq}}(2,2)}\otimes\ket{00}$, the smallest recorded pruned solution contains fourteen optical edges and reaches unit fidelity with the ancilla restriction, while the unrestricted solution records thirteen edges. This showcases potential use cases for \texttt{Grinch} in applying complex unitary operations, as one can search for the representation that encodes the desired transformation.

In general, the tabulated gate fidelities quantify agreement with the normalized encoded input--output target in the ideal model. They are not experimental gate fidelities or probabilities of successful operation. However, our framework offers a flexible reward function that can be modified to incorporate additional requirements. This reward-based formulation could be extended to include hardware-dependent metrics, provided that a suitable model or evaluator for those quantities is supplied; such metrics are not included in the present fidelity-only calculations. Additionally, as we have shown, hardware restrictions can be implemented through a reduced set of actions in addition to having reward penalties for final states that violate such conditions.

\section{Discussion}
The closest generative precedent to \texttt{Grinch} is the quantum-optics variational autoencoder (QOVAE) \cite{QOVAE}, which learns a continuous latent representation of optical-device sequences and can generate experiments that resemble selected distributions in its training data. This data-driven formulation has several practical consequences: a representative collection of experiments must first be generated, the learned sampling distribution is tied to that collection, and targeting an individual state requires a separate search engine based on the latent space representation. The QOVAE study also notes that incorporating continuously parametrized devices requires discretization and that generated experiments may remain challenging to implement in the laboratory. \texttt{Grinch} provides a complementary approach to these limitations by learning directly from a state-dependent reward rather than reconstructing a distribution over archived experiments. Budget-aware perfect-matching masks restrict graph construction, while the GFlowNet objective learns to distribute probability among high-reward terminal graphs. Continuous edge-weight optimization evaluates agreement with the target state for eligible terminal candidates. The structural masks reduce the search space, improving efficiency, while action-set restriction can impose hardware constraints directly.

Consequently, \texttt{Grinch} can perform target-directed and diversity-seeking circuit discovery without a pre-existing dataset.
\texttt{Grinch}'s bottom-up approach is particularly advantageous when ancilla photons are needed, as it allows us to expand the mapping between optical graphs and logical gates, such as CNOT and Toffoli gates. This opens the possibility of extending the framework beyond individual gate discovery toward the generation of more complex quantum circuits.
In top-down-based approaches, each added ancilla enlarges the initial dense graph and the number of edges that must be pruned. In contrast, in \texttt{Grinch}, an ancilla only adds $n n_a d d_a$ possible actions to the total, as shown in Eq. \ref{eq:transformer_action_dimension}, keeping the sampled graph sparse. This makes searching for an odd number of photons in an entangled state more efficient than other top-down frameworks.

The SRV solutions found with additional ancillas also illustrate how the sampled graphs can guide the development of an experimental design. A high numerical fidelity alone leaves open the question of whether the proposed setup can provide useful count rates. The analytic constructions in Appendix Sec.~\ref{sec:srv-constructions} address this question by identifying the relevant interference mechanisms and providing tunable weight assignments on each discovered topology. Once source and detection parameters are specified, these families allow different operating points to be assessed against the fidelity and count-rate requirements of the experiment. This also supports the use of \texttt{Grinch} for interpretable scientific discovery: its sampled topologies provide a basis for extracting interference rules and constructing parametrized families of optical experiments.

Sequential, constructive generation also appears in recent language-model-based approaches to quantum-experiment design. The meta-design framework uses an autoregressive transformer to generate executable programs token by token, with the resulting programs encoding construction rules for entire families of quantum experiments \cite{ArltMetaDesign}. AI-Mandel similarly uses LLM agents to iteratively formulate and refine scientific ideas and executable instructions \cite{ArltAIMandel}. However, these methods are not bottom-up circuit-search approaches in the same sense as \texttt{Grinch}: AI-Mandel delegates circuit discovery to PyTheus, whose topological optimization proceeds from a large initial graph through edge removal \cite{PyTheus}, whereas meta-design \cite{ArltMetaDesign} generates programs rather than circuit graphs and relies on a large synthetic training set. In contrast, \texttt{Grinch} applies sequential construction directly to the experimental topology, adding edges under physical constraints and learning from a task-specific reward without prior data. The common principle is therefore constructive generation, while the representation, supervision, and level at which construction occurs are fundamentally different. Thus, \texttt{Grinch} and the resulting trained models become an attractive alternative for agentic frameworks as tools for generating quantum optics experiments. 

\section{Summary}
We introduced \texttt{Grinch}, a GFlowNet-based framework that samples quantum optical circuits in proportion to a task-defined reward. Unlike conventional generative models that are trained on existing datasets, \texttt{Grinch} learns directly from reward evaluations, enabling circuit discovery when representative training data is unavailable. By combining masked, edge-by-edge graph construction with continuous edge-weight optimization, the framework addresses coupled discrete--continuous design problems while restricting exploration through structural masks.

Across multipartite-entangled-state, graph- and cluster-state, and nonlocal photonic-gate benchmarks, \texttt{Grinch} identified multiple high-fidelity optical-graph solutions. Modest flexibility beyond the strict terminal edge budget accelerated the discovery of high-reward circuits, after which pruning removed redundant edges while maintaining fidelity. Moreover, when alternative implementations exist, the sampled solutions occupy multiple high-reward regions of graph space, demonstrating that \texttt{Grinch} can recover diverse circuit realizations rather than a single optimum. By constructing graphs using the bottom-up implementation, \texttt{Grinch} handles ancilla photons naturally: ancilla adds only a modest number of candidate edges rather than enlarging a dense graph to be pruned, which enables the multi-ancilla constructions for asymptotic states and Toffoli gates reported here. Importantly, \texttt{Grinch} can identify asymptotic solutions for states that cannot be exactly generated by graphs and derive alternative arrays for nonlocal quantum gates under hardware constraints, as demonstrated in the no-ancilla communication experiment reported here.

The broader applicability of this approach depends on a sufficiently efficient simulator or optimizer to evaluate terminal rewards, since these evaluations are performed repeatedly during training. Building on the state-preparation and nonlocal photonic-gate applications considered here, future extensions could address additional optical architectures, hardware-dependent objectives, and inclusion of \texttt{Grinch} as a domain-specific tool within emerging agentic frameworks for quantum experiment design \cite{nonLocal-agents,DesignPhysExpAINature2026}. 

\section*{Data availability}
\texttt{Grinch} is available at \href{https://github.com/ishume94/Grinch}{https://github.com/ishume94/Grinch}. This repository contains the necessary code to reproduce the results reported in this work.

\section*{Acknowledgments}
ILHM acknowledges support from Sandbox AQ's 2025 Research Excellence Scholarship.
This research was partly enabled by support from the Digital Research Alliance of Canada and NSERC Discovery Grant No. RGPIN-2024-06594. 

\appendix
\section{GFlowNets} \label{sec:app-gflownets}
Generative Flow Networks (GFlowNets) \cite{bengio2021flow} are a type of generative model that allows learning the probability of sampling a compositional object $s \in \mathcal{S}$ through a sequence of stochastic steps employing a model to make the probability of reaching it proportional to a positive reward function $R(s)$,
\begin{equation}
    P_\mathcal{T}(s)\propto R(s).
\end{equation}
The object construction can be mapped to a directed acyclic graph (DAG) that generates the objects that we wish to sample through a set of discrete actions $\mathcal{A}$; this is $G=(\mathcal{S},\mathcal{A})$. The set of actions is comprised of the transitions between two states $s\rightarrow s'$ with a trajectory $\tau\in\mathcal{T}$ defined by the sequence $\tau=(s_0\rightarrow s_1\rightarrow\dots s_{f-1}\rightarrow s_f)$ where $s_0$ and $s_f$ represent initial and terminating states, respectively. 

GFowNets requires a function that parametrizes the probability of reaching each terminating state from the set of all possible trajectories. We refer to this function as a \textit{flow}, denoted \( F \), which is any non-negative function defined over the set of trajectories \( \mathcal{T} \), as \( F : \mathcal{T} \mapsto \mathbb{R}^+ \). A defining characteristic of GFlowNets is imposing Markovian flows, which requires that for any state \( s\neq s_0 \), outgoing edge \( s\rightarrow s'\), and for any partial trajectory \( \tau = (s_0,s_1,\dots, s_n=s) \in \mathcal{T}^{partial}\)
\begin{equation}
    P(s\rightarrow s'|\tau) = P(s\rightarrow s'|s) = P_F (s'|s).
\end{equation}
Markovian Flows have the property that the probabilities of complete trajectories factorize according to the DAG; this is for any complete trajectory \(\tau = (s_0,s_1\dots,s_{n}=s_f)\):
\begin{align}
    &P(\tau) = \prod_{t=1}^{n}P_F(s_{t}|s_{t-1}) \label{eqn:trajF} \\
    &P(\tau|s_n=s_f)=\prod_{t=1}^{n}P_B(s_{t-1}|s_t) \label{eqn:trajB}
\end{align}

The Markovian flow enables the computation of transition probabilities between states. Specifically, the forward transition probability from a state \( s \) to a child state \( s' \) is defined as:
\begin{equation}
    \forall (s \rightarrow s') \in \mathcal{A}, \quad P_F(s'|s) := P(s \rightarrow s' | s) = \frac{F(s \rightarrow s')}{F(s)}. \label{eqn:Forward}
\end{equation}
Similarly, the backwards transition probability from a state \(s'\) to a parent state \(s\) is:
\begin{equation}
    \forall (s \rightarrow s') \in \mathcal{A}, \quad P_B(s|s') := P(s \rightarrow s' | s') = \frac{F(s \rightarrow s')}{F(s')}.  \label{eqn:Backward}
\end{equation}
Additionally, the probability of reaching a terminal state \( s_f \in \mathcal{S}^f \) is given by:
\begin{equation}
    \forall s_f \in \mathcal{S}^f, \quad P_T(s_f) := \frac{F(s \rightarrow s_f)}{Z}
\end{equation}
with $Z$ being the partition function defined as $ Z := F(\mathcal{T}) = \sum_{\tau \in \mathcal{T}} F(\tau)$ which represents the total flow for all complete trajectories.

The Markovian nature of the flows leads to different training objectives for GFlowNets. The Flow Matching conditions imply that for any state that is not initial or terminal, the total incoming flow equals the total outgoing flow \cite{bengio2021flow}:
\begin{equation}
F(s) = \sum_{s' \in \text{ parents}(s)} F(s' \rightarrow s) = \sum_{s'' \in \text{ children}(s)} F(s \rightarrow s''). 
\end{equation}
Combining the relations between Forward and Backward transition probabilities (Eqs. \ref{eqn:Forward},\ref{eqn:Backward}) leads to the detailed balance constraint \cite{bengio2023gflownet}:
\begin{equation}
    F(s)P_F(s'|s)=F(s')P_B(s|s').
\end{equation}
Finally, through the trajectory decomposition from Eqs. \ref{eqn:trajF}-\ref{eqn:trajB} with the fact that \(P(\tau|s_n=s_f)=P(\tau)/P(s_f)\) leads to the trajectory balance condition \cite{malkin2022trajectory}:
\begin{equation}
    Z \prod_{t=1}^{n}P_F(s_{t}|s_{t-1})=F(s_f)\prod_{t=1}^{n}P_B(s_{t-1}|s_t) 
\end{equation}
where the flow through the final state is set equal to its reward function for training purposes \(F(s_f) = R(s_f)\)
GFlowNet parametrizes these flow functions as $F_\theta(s',s)$ with a set of learnable parameters $\theta$ which can be trained to minimize a loss $\mathcal{L}$ function coming from either of the aforementioned training objectives. The GFlowNet is then the tuple $\text{GFlow}=(G,R,F_\theta)$ with $R$ a reward function and $F_\theta$ a flow parametrization of our DAG \cite{zhang2023letFlows}. The training protocol is carried out concurrently with the sampling stage. We begin with equal probabilities for object construction and, while we reach terminating states, the GFlowNet model updates the probabilities according to a reward function by minimizing the loss function. For this work, we consider the trajectory balance loss \cite{malkin2022trajectory} function, which gives increased credit assignment, making the training protocol more efficient. The trajectory balance loss is defined as,
\begin{equation}
    \mathcal{L}_{TB}(\tau)=\left(\log\frac{Z_\theta\prod_{t=1}^nP_\theta(s_t|s_{t-1})}{R(s_n)\prod_{t=1}^nP_\theta(s_{t-1}|s_t)} \right)^2 \label{eq:lossTB_}
\end{equation}
with $s_n\in\mathcal{S}^f$. 

GFlowNets thus offer a principled method for sampling from a distribution proportional to a reward function \( R(s) \), or equivalently from an energy function \( E(s) := -\log R(s) \) \cite{bengio2023gflownet}. These models are particularly effective in problems exhibiting a compositional structure, where sequential generation is natural, the reward function is non-negative and easy to define, and the target distribution is highly multimodal \cite{jain2023gflownets}. GFlowNets has been used for several applications, among them solving graph combinatorial problems \cite{zhang2023letFlows}, replacing Bayesian optimization and reinforcement learning methods to find protein sequences in antimicrobials and DNA sequences \cite{bioseq}, and molecule generation for target properties \cite{jain2023gflownets}. 

\section{GFlowNets for Quantum Optical Circuits}\label{sec:app-gflow-qoc}

For the design of quantum experiments, the GFlowNets framework leverages diverse graph generation via a sampling procedure. The initial state $s_0$ is an edgeless graph on the fixed set of $N=n+n_a$ nodes, including ancillas $n_a$, each representing a photonic path, with vertices $V=\{v_1,\dots, v_N\}$. The set of actions, $\mathcal{A}$, is given by the addition of a bicolored weighted edge $E(w)$, with the weight defined by the mode numbers ($k$) and vertices as $w_{v_i,v_j}^{k_i,k_j}$. The maximum number of modes gives the maximum number of colors available for the graph ($d$).

GFlowNets is responsible for generating diverse experimental set-ups; however, the resulting graph requires an optimization procedure over the continuous weights to yield the final fidelity. Top-down approaches require generating an initial graph with $|E_G|=d^2[n(n-1)/2]$ edges, followed by minimizing a loss function. After convergence is achieved, a topological optimization step follows: an edge is removed from the graph, and the minimization is repeated, iteratively reducing the graph's size. Our approach leverages the fact that minimizing the loss function is computationally efficient, enabling us to construct a reward function to train the GFlowNet. This helps to produce amortized sampling, avoiding unnecessary minimizations and reducing computational resources.

The reward function employed uses the fidelity $\F$ of the target state for the given experiment as
\begin{equation}
    R(s)= \F^2(w_s).
\end{equation}
 To find the optimal values $w_s$, the following minimization is performed for every terminal state
\begin{equation}
    w_s(s_f)=\argmin_{w_{s_f}} \left[-\F(w_{s_f})\right]. \label{eq:weight_opt}
\end{equation}
An additional $L_1$ regularization term $ + \gamma|w_{s_f}|_1$ in the optimization was tested for states that cannot be prepared perfectly with linear optics and probabilistic photo-pair sources, like $\ket{GHZ(6,3)^*}$, with $\gamma$ set to 0.01, similarly to the loss minimization from \textsc{Theseus} \cite{TheseusDesign}. We did not observe any improvement in our method's performance with the $L_1$ term; thus, we keep the standard protocol in Eq. \ref{eq:weight_opt} across simulations. The optimal weights and corresponding rewards are stored to avoid recalculating them during sampling, and a function is employed to assess whether two graphs are equivalent. 

The reward is then used to learn the flows between states on the DAG by minimizing the trajectory-balance loss function in Eq. \ref{eq:lossTB}, with the flow model implemented as a neural network with parameters $\theta$. This optimization is done through stochastic gradient descent with the ADAM optimizer. 
If the found fidelity satisfies $\F \geq 0.95$, we employ a pruning procedure based on logical clauses to try to find a solution with a lower number of edges.

\section{Training and architecture} \label{sec:training_and_architecture}
We parametrize the forward and backward policies, $P_F$ and $P_B$, using a shared graph-transformer encoder followed by a multilayer perceptron (MLP). The encoder uses the edge-aware \texttt{TransformerConv} operator implemented in PyTorch Geometric, which performs multihead dot-product attention over graph neighborhoods \cite{GraphTransformerPyTorch,GeneralizationTransformerOnGraphs}. We use three graph-transformer layers with four attention heads per layer and set \texttt{n\_hid\_units}$=128$ for all calculations. The hidden node representations and the pooled graph representation therefore have dimension 128.

\paragraph{Graph representation and action space.}
Let $n$ denote the number of non-ancilla nodes, $n_a$ the number of ancilla nodes, and $N=n+n_a$ the total number of nodes. A partial graph $s_t$ contains $E_t$ selected edges. Each node is represented by an $N$-dimensional one-hot vector identifying its optical path, so the node-feature matrix is $X=I_N\in\mathbb{R}^{N\times N}$. All $N$ nodes are included, even when they are isolated. Each selected edge $((u,v),(c_u,c_v))$ has the two-dimensional attribute vector $(c_u,c_v)$, containing its endpoint colors as numerical labels. Thus, \texttt{node\_feat\_dim}$=N$, \texttt{edge\_feat\_dim}$=2$, and the tensors \texttt{edge\_index} and \texttt{edge\_attr} have shapes $2\times E_t$ and $E_t\times 2$, respectively. The graph encoding stores each edge once, in the direction $u\rightarrow v$ with $u<v$; it does not add the reverse edge. Consequently, attention follows these directed edges rather than a bidirectional or fully connected attention graph. Optimized optical edge weights are not included in the neural network inputs.

The ordered list \texttt{FEATURE\_KEYS} enumerates the candidate edge additions, with each entry specifying a node pair and an ordered pair of endpoint colors. We denote its length by $|\mathcal{A}|=\operatorname{len}(\texttt{FEATURE\_KEYS})$. Distinct color assignments to the same node pair correspond to distinct actions. For state preparation with $d$ colors per non-ancilla node and $d_a$ allowed colors per ancilla node, the implemented action-space construction gives
\begin{equation}
|\mathcal{A}|=\binom{n}{2}d^2+n n_a d d_a.
\label{eq:transformer_action_dimension}
\end{equation}
The default ancilla color is fixed to zero, corresponding to $d_a=1$; otherwise, $d_a$ can be set by \texttt{c\_a} in the accompanying library. For quantum-gate targets, we remove input--input edges before constructing the model. Accordingly, $|\mathcal{A}|$ determines the output dimension of each policy. Subsequent action masking changes which entries can be selected without changing the size of either policy output. For example, $n=4$, $d=2$, and $n_a=0$ give $|\mathcal{A}|=24$, so the decoder produces 48 logits, split into two vectors of length 24.

\paragraph{Graph-transformer encoder and policy decoder.}
We denote the hidden dimension by $d_{\mathrm{hid}}$ (\texttt{hidden\_dim}, set through \texttt{n\_hid\_units} in the included library), the number of transformer layers by $L$ (\texttt{num\_layers}), and the number of attention heads by $H$ (\texttt{heads}). We use $d_{\mathrm{hid}}=128$, $L=3$, $H=4$, and \texttt{dropout=0.0} for all calculations.

An initial linear layer maps each node feature from $N$ dimensions (\texttt{node\_feat\_dim}$=N$) to $d_{\mathrm{hid}}$ dimensions. Each of the $L$ transformer layers has a separate edge encoder consisting of an MLP with dimensions $2\rightarrow d_{\mathrm{hid}}\rightarrow d_{\mathrm{hid}}$ and a ReLU activation between its linear layers. Here, \texttt{edge\_feat\_dim}$=2$ corresponds to the two endpoint-color attributes. These edge encoders act on the original two-component color attributes at every layer. Within \texttt{TransformerConv}, the encoded edge attributes are further projected and added to both the attention keys and the message values.

Each attention head has $d_{\mathrm{hid}}$ channels. Thus, the internal query, key, and value projections each have $H d_{\mathrm{hid}}$ channels per node, reshaped into $H$ heads of dimension $d_{\mathrm{hid}}$. For the values used in our calculations, each projection therefore has 512 channels, organized into four heads with 128 channels each. The $H$ head outputs are averaged, leaving a $d_{\mathrm{hid}}$-dimensional node representation. Each convolution also uses a learned gate, enabled by \texttt{beta=True}, to combine the aggregated message with a linearly transformed root-node representation. A ReLU activation and an additional residual connection follow each convolution. The layer update can be written as
\begin{equation}
\mathbf{h}_i^{(\ell+1)}=\mathbf{h}_i^{(\ell)}+\operatorname{ReLU}\!\left(\mathbf{t}_i^{(\ell)}\right),\qquad \ell=0,\ldots,L-1,
\label{eq:transformer_residual_update}
\end{equation}
where $\mathbf{h}_i^{(0)}$ is the initial $d_{\mathrm{hid}}$-dimensional node embedding and $\mathbf{t}_i^{(\ell)}$ is the gated output of the corresponding edge-aware transformer convolution. No layer normalization or separate node-wise feedforward sublayer is used.

A final linear transformation with dimensions $d_{\mathrm{hid}}\rightarrow d_{\mathrm{hid}}$ produces the node embeddings $\mathbf{z}_i$. Global sum pooling then gives the graph embedding
\begin{equation}
\mathbf{g}(s_t)=\sum_{i=1}^{N}\mathbf{z}_i\in\mathbb{R}^{d_{\mathrm{hid}}}.
\label{eq:transformer_graph_pooling}
\end{equation}

The decoder is an MLP with dimensions $d_{\mathrm{hid}}\rightarrow d_{\mathrm{hid}}\rightarrow2|\mathcal{A}|$, with a ReLU activation between its linear layers. Thus, \texttt{hidden\_dim} determines the size of the hidden representations, whereas the length of \texttt{FEATURE\_KEYS} determines the policy output dimensions. The decoder output is split into two vectors of unnormalized logits,
\begin{equation}
\bigl[\boldsymbol{\ell}_F(s_t),\boldsymbol{\ell}_B(s_t)\bigr]=W_2\operatorname{ReLU}\!\left(W_1\mathbf{g}(s_t)+\mathbf{b}_1\right)+\mathbf{b}_2,
\label{eq:transformer_policy_logits}
\end{equation}
where $W_1\in\mathbb{R}^{d_{\mathrm{hid}}\times d_{\mathrm{hid}}}$, $W_2\in\mathbb{R}^{2|\mathcal{A}|\times d_{\mathrm{hid}}}$, $\mathbf{b}_1\in\mathbb{R}^{d_{\mathrm{hid}}}$, and $\mathbf{b}_2\in\mathbb{R}^{2|\mathcal{A}|}$. Each logit vector has dimension $|\mathcal{A}|$, with one entry per candidate action. The final decoder layer contains $2|\mathcal{A}|(d_{\mathrm{hid}}+1)$ trainable parameters, including its biases, corresponding to $258|\mathcal{A}|$ parameters for $d_{\mathrm{hid}}=128$. For a batch of $B$ graphs, the pooled graph embeddings have shape $B\times d_{\mathrm{hid}}$, the decoder output has shape $B\times2|\mathcal{A}|$, and each policy-logit tensor has shape $B\times |\mathcal{A}|$ before singleton dimensions are removed.

 \begin{lstlisting}[language=Python, caption={Graph-transformer model used to parametrize the forward and backward policies. The forward pass returns unnormalized action logits. The scalar \texttt{logZ} is learned jointly with the policy parameters.}]
class GraphTransformerEncoder(nn.Module):
    """
    Graph Transformer encoder using PyG's TransformerConv (attention on graphs).
    Supports edge features via edge_dim.
    """
    def __init__(
        self,
        node_feat_dim: int,
        edge_feat_dim: int,
        hidden_dim: int,
        num_layers: int = 3,
        heads: int = 4,
        dropout: float = 0.0,
        use_residual: bool = True,
    ):
        super().__init__()
        self.node_encoder = nn.Linear(node_feat_dim, hidden_dim)

        self.edge_encoders = nn.ModuleList()
        self.convs = nn.ModuleList()
        self.dropout = float(dropout)
        self.use_residual = bool(use_residual)

        for _ in range(num_layers):
            self.edge_encoders.append(
                nn.Sequential(
                    nn.Linear(edge_feat_dim, hidden_dim),
                    nn.ReLU(),
                    nn.Linear(hidden_dim, hidden_dim),
                )
            )
            # concat=False keeps output dim = hidden_dim regardless of heads
            self.convs.append(
                TransformerConv(
                    in_channels=hidden_dim,
                    out_channels=hidden_dim,
                    heads=heads,
                    concat=False,
                    dropout=dropout,
                    edge_dim=hidden_dim,
                    beta=True,  # learnable skip connection gate
                )
            )

        self.final = nn.Linear(hidden_dim, hidden_dim)

    def forward(self, x, edge_index, edge_attr):
        x = self.node_encoder(x)

        for conv, edge_encoder in zip(self.convs, self.edge_encoders):
            h_in = x
            e = edge_encoder(edge_attr) if edge_attr is not None else None

            x = conv(x, edge_index, e)
            x = nn.functional.relu(x)
            x = nn.functional.dropout(x, p=self.dropout, training=self.training)

            if self.use_residual:
                x = x + h_in

        return self.final(x)
        
class Transformer_TBModel(nn.Module):
    def __init__(
        self,
        node_feat_dim: int,
        edge_feat_dim: int,
        hidden_dim: int,
        FEATURE_KEYS,
        num_layers: int = 3,
        heads: int = 4,
        dropout: float = 0.0,
    ):
        super().__init__()
        self.encoder = GraphTransformerEncoder(
            node_feat_dim=node_feat_dim,
            edge_feat_dim=edge_feat_dim,
            hidden_dim=hidden_dim,
            num_layers=num_layers,
            heads=heads,
            dropout=dropout,
        )
        self.pool = global_add_pool
        self.decoder = nn.Sequential(
            nn.Linear(hidden_dim, hidden_dim),
            nn.ReLU(),
            nn.Linear(hidden_dim, 2 * len(FEATURE_KEYS)),  # Forward and Backward
        )
        self.logZ = nn.Parameter(torch.ones(1))
        self.FEATURE_KEYS = FEATURE_KEYS

    def forward(self, data):
        x, edge_index, edge_attr, batch = data.x, data.edge_index, data.edge_attr, data.batch
        node_emb = self.encoder(x, edge_index, edge_attr)
        graph_emb = self.pool(node_emb, batch)
        logits = self.decoder(graph_emb)
        P_F = logits[..., : len(self.FEATURE_KEYS)]
        P_B = logits[..., len(self.FEATURE_KEYS) :]
        return P_F.squeeze(), P_B.squeeze()
\end{lstlisting}

\paragraph{Policy normalization and training.}
The logits returned by the model are converted into categorical distributions after applying the masking and progress rules described in Sec.~\ref{sec:masking_and_pruning}. The forward policy excludes duplicate and structurally inadmissible edges, applies the remaining-budget completion bounds, and adds a finite-progress preference to the admitted logits. Excluded logits are set to $-\infty$ before normalization by \texttt{Categorical(logits=...)}. The backward mask allows the removal of any edge present in the current graph, without additional progress preference. A forward action is sampled from the resulting effective policy, and its backward log probability is evaluated at the successor state. These effective policy probabilities are used in the trajectory-balance loss.

Each trajectory starts from the edgeless graph on the fixed set of $N$ nodes and terminates upon reaching $T=\texttt{max\_edges}$ edges or exhausting its admissible forward actions. There is no learned termination action. Upon early action exhaustion, a graph containing a perfect matching for every required target component receives the usual optimized fidelity-based reward; otherwise, its reward is zero. The reward entering the logarithm in the trajectory-balance loss is clipped below at $10^{-30}$. We denote the terminal graph by $s_f$, since its number of edges need not equal $T$.
The terminal graph is evaluated by a separate continuous optimization of its real edge weights using L-BFGS-B with bounds $[-1,1]$. The scalar returned by this reward calculation is $\mathcal{R}(s_T)$ and enters the trajectory-balance loss shown in Eq. \ref{eq:lossTB}. Here, $\log Z$ is a single trainable scalar shared across all states within a training run, initialized to 1 and optimized jointly with the policy parameters. For all calculations, we used a learning rate of $10^{-3}$ for the policy parameters and $10^{-2}$ for $\log Z$. Trajectory losses are accumulated by summation between optimizer updates, collecting five trajectories per update in all cases.

\section{Masking and pruning logic}
\label{sec:masking_and_pruning}

Masking restricts the edge additions available during graph construction, whereas pruning attempts to simplify a completed graph after its edge weights have been optimized. Both procedures use perfect matchings on the full set of $N=n+n_a$ nodes, including ancillas. Their roles differ: the forward mask applies structural completion bounds within the edge budget, while pruning combines a target-support condition with a numerical fidelity threshold. In contrast to Klaus \cite{Klaus}, \texttt{Grinch} evaluates these conditions using matching computations rather than by calling a general-purpose Boolean satisfiability solver.

\paragraph{Target support and candidate edges.}
Let the normalized target, including any ancillas, be
\begin{equation}
\ket{\psi}_{\mathrm{Target}}=\sum_{b\in\mathcal{B}_{\mathrm{tar}}}t_b\ket{b},\qquad \sum_{b\in\mathcal{B}_{\mathrm{tar}}}|t_b|^2=1,
\label{eq:mask_target_support}
\end{equation}
where $b=(b_1,\ldots,b_N)$ specifies the mode at every node and $\mathcal{B}_{\mathrm{tar}}$ contains the basis strings with nonzero target amplitudes. During training, this support is extracted from \texttt{target\_state}. The masks depend on which coefficients are nonzero, but not on their relative magnitudes or phases; these enter through the fidelity evaluation.

The action set $\mathcal{A}$ is the fixed list \texttt{FEATURE\_KEYS} defined in Sec.~\ref{sec:training_and_architecture}. Each action $a$ adds a colored edge $e_a=((u,v),(c_u,c_v))$. Candidate edges already respect the configured ancilla colors and, in quantum-gate mode, exclude input--input edges. The supplied action-space construction contains no self-loops and can be modified to remove ancilla--ancilla edges. Write $e_a\sim b$ when $(c_u,c_v)=(b_u,b_v)$. An edge passes the initial target-compatibility filter only if $e_a\sim b$ for at least one $b\in\mathcal{B}_{\mathrm{tar}}$. A further filter removes edges that cannot participate in any perfect matching covering all $N$ nodes within this pair-compatible candidate graph. This second test ignores the output coloring of the complete matching: an admitted edge need not belong to a matching that generates a target basis string. Denote the surviving colored-edge set by $\mathcal{E}_{\mathrm{adm}}$. These filters restrict the search space without changing $|\mathcal{A}|$ or the decoder dimensions.

\paragraph{Budget-aware forward masking.}
For each target basis string $b$, let $\mathcal{P}_b$ be the family of perfect matchings drawn from $\mathcal{E}_{\mathrm{adm}}$ whose endpoint colors produce $b$. The completion bounds below apply to target--action-space combinations for which every $\mathcal{P}_b$ is nonempty. For a partial graph $s$ with edge set $E(s)$, define
\begin{equation}
\delta_b(s)=\min_{P\in\mathcal{P}_b}|P\setminus E(s)|.
\label{eq:matching_completion_cost}
\end{equation}
Thus, $\delta_b(s)$ is the minimum number of additional edges required to obtain at least one matching for $b$, and $\delta_b(s)=0$ means that $s$ already contains such a matching. \texttt{Grinch} evaluates these costs by a memoized recursion over subsets of unmatched nodes, assigning cost zero to an edge already present and cost one to an absent candidate edge. Completion costs are cached across repeated partial graphs. This calculation does not optimize the continuous edge weights or enumerate all possible completed circuit graphs.

Let $T=\texttt{max\_edges}$, let $s_a$ be the graph obtained by adding $e_a$ to $s$, and let $r=T-|E(s)|-1$ be the remaining budget after that addition. Define
\begin{equation}
\mu=\max_{e\in\mathcal{E}_{\mathrm{adm}}}\sum_{b\in\mathcal{B}_{\mathrm{tar}}}\mathbf{1}[e\sim b],
\label{eq:matching_basis_incidence}
\end{equation}
where $\mathbf{1}[\cdot]$ is an indicator function; with $\mu=1$ when the admitted edge set is empty. The forward mask is
\begin{equation}
\begin{aligned}
\mathcal{M}(s,a)={}&\mathbf{1}[e_a\in\mathcal{E}_{\mathrm{adm}}\setminus E(s)]\\
&\times\mathbf{1}[\delta_b(s_a)\leq r\ \text{for every }b\in\mathcal{B}_{\mathrm{tar}}]\\
&\times\mathbf{1}\!\left[\sum_{b\in\mathcal{B}_{\mathrm{tar}}}\delta_b(s_a)\leq\mu r\right].
\end{aligned}
\label{eq:matching_forward_mask}
\end{equation}
The first condition excludes duplicate and statically inadmissible edges. The second requires each target component to remain individually reachable within the remaining budget. The third couples these requirements through a shared-budget bound: one added edge can reduce the completion cost by at most one for each of at most $\mu$ target components. These are necessary conditions for a joint completion, not sufficient conditions. In particular, the separately optimal completions for different target components need not be simultaneously achievable within the budget. Structural support also does not guarantee the target amplitudes or a target fidelity.

\paragraph{Progress preference and policy normalization.}
In addition to the hard mask, \texttt{Grinch} adds a finite progress preference to the forward logits. Let $\mathcal{U}(s)=\{b\in\mathcal{B}_{\mathrm{tar}}:\delta_b(s)>0\}$ be the currently unsupported target components. For an admitted action, define
\begin{equation}
g(s,a)=\begin{cases}
\displaystyle\frac{1}{|\mathcal{U}(s)|}\sum_{b\in\mathcal{U}(s)}\left[\delta_b(s)-\delta_b(s_a)\right],&|\mathcal{U}(s)|>0,\\[6pt]
0,&|\mathcal{U}(s)|=0.
\end{cases}
\label{eq:matching_progress_preference}
\end{equation}
A single edge changes each completion cost by at most one, so $0\leq g(s,a)\leq1$. The coefficient multiplying this preference is one in \texttt{Grinch}. The effective forward logits are
\begin{equation}
\widetilde{\ell}_F(s,a)=\begin{cases}
\ell_F(s,a)+g(s,a),&\mathcal{M}(s,a)=1,\\
-\infty,&\mathcal{M}(s,a)=0,
\end{cases}
\label{eq:matching_masked_forward_logits}
\end{equation}
and $P_F$ is the categorical distribution obtained by applying softmax to these effective logits whenever at least one action is admitted. The same effective distribution supplies the forward log probabilities in the trajectory-balance loss. This preference favors progress toward missing target components without imposing a hard requirement of immediate progress. Preparatory edges and additional matchings for already-supported components remain selectable when they pass Eq.~\ref{eq:matching_forward_mask}.

The backward mask admits removal of any edge currently present:
\begin{equation}
\mathcal{M}_B(s,a)=\mathbf{1}[e_a\in E(s)].
\label{eq:matching_backward_mask}
\end{equation}
Excluded backward logits are also set to $-\infty$, and no progress preference is added. The backward probability associated with a sampled addition is evaluated at the successor graph. For forward-reachable graphs, removing one edge increases each completion cost by at most one and their sum by at most $\mu$, while restoring one unit of budget. This is consistent with the forward completion bounds. Unlike pruning, a backward transition need not preserve the target support already present in the child graph, because its parent is a partial construction.

During sampling, a trajectory ends when it reaches $T$ edges or when its forward mask becomes empty; hence its length is at most $T$. If an exhausted graph contains a matching for every target component, its weights are optimized, and its reward is evaluated normally. Otherwise, it receives zero reward, with the reward argument of the logarithm clipped below at $10^{-30}$ in the trajectory-balance loss. Target support alone does not terminate a trajectory while admitted actions and budget remain. There is no learned termination action, and repeated edges are not added to fill an unused budget.

\paragraph{Logical support condition for pruning.}
Pruning uses the numerical support set
\begin{equation}
\mathcal{B}_{\mathrm{pr}}=\{b:|t_b|>\varepsilon_{\mathrm{supp}}\},\qquad \varepsilon_{\mathrm{supp}}=10^{-12},
\label{eq:pruning_target_support}
\end{equation}
corresponding to \texttt{support\_tol} in the library. For the graph submitted to pruning, all perfect matchings on the full target node set are enumerated once, retaining their edge indices and output basis strings. Let $z_e\in\{0,1\}$ indicate whether an edge of that graph is retained. The target-support condition, denoted as the S clause in \texttt{Grinch}, can be written as
\begin{equation}
S(\mathbf{z})=\bigwedge_{b\in\mathcal{B}_{\mathrm{pr}}}\left[\bigvee_{P\in\mathcal{P}_b(s_f)}\bigwedge_{e\in P}z_e\right],
\label{eq:pruning_support_clause}
\end{equation}
where $\mathcal{P}_b(s_f)$ contains the perfect matchings of the submitted terminal graph that produce $b$. A proposed deletion is rejected immediately if it removes the last surviving matching for any required target component. Subsequent support tests reuse the initial matching data by discarding matchings containing deleted edges. No new matching enumeration is needed after each deletion.

The unwanted-output condition called the C clause in Klaus \cite{Klaus}, which would reject a nontarget basis string generated by exactly one perfect matching, is not considered here, as intruder perfect matchings can be removed asymptotically through the weight optimization. The amplitude of any generated basis string is determined by
\begin{equation}
A_b(\mathcal{G},\vw)=\sum_{P\in\mathcal{P}_b(\mathcal{G})}\prod_{e\in P}w_e.
\label{eq:pruning_matching_amplitude}
\end{equation}
Consequently, counting matchings alone does not determine the normalized output state: unwanted amplitudes can be suppressed through weight optimization, and multiple pathways can interfere. Such outputs are not excluded by the S clause, but their amplitudes are included in the numerical fidelity calculation, including its normalization. The procedure therefore separates structural support from agreement with the full target state.

\paragraph{Greedy deletion with a fidelity threshold.}
Pruning is attempted when the optimized weights are available, and the unpruned reward satisfies
\begin{equation}
\mathcal{R}(s_f)=\mathcal{F}(s_f,\vw)^2\geq\mathcal{F}_{\mathrm{pr}}^2,\qquad \mathcal{F}_{\mathrm{pr}}=0.95.
\label{eq:pruning_entry_threshold}
\end{equation}

The pruning utility first removes repeated copies of an identical colored edge, retaining its first occurrence and corresponding weight; distinct color pairs on the same node pair remain distinct edges. After this canonicalization, no further pruning or weight optimization is attempted if the graph does not cover the full target node set, fails the S clause, or has nonfinite fidelity or fidelity below $\mathcal{F}_{\mathrm{pr}}$ at its supplied weights.

At the beginning of each greedy sweep, the retained edges are ordered by increasing current $|w_e|$. For each edge, the routine first checks whether its deletion preserves Eq.~\ref{eq:pruning_support_clause}. If it does, the fidelity is evaluated with the surviving weights. The deletion is accepted immediately when this fidelity is finite and at least $\mathcal{F}_{\mathrm{pr}}$. Otherwise, the surviving weights initialize an L-BFGS-B optimization of $-\mathcal{F}$ with real bounds $w_e\in[-1,1]$. The resulting fidelity is recomputed, and the deletion is accepted only if it meets the same threshold. A failed test leaves the previous graph and weights unchanged. 

Accepted deletions update the retained graph, its weights, and its surviving matching set. Sweeps continue until a complete sweep removes no edge or 10 passes are reached. The routine returns the retained graph, its aligned weights, and a Boolean keep-mask over the original edge list. Pruning enforces a fidelity floor, not equality with the unpruned fidelity: an accepted deletion may reduce fidelity provided that $\mathcal{F}\geq\mathcal{F}_{\mathrm{pr}}$. Accepted outputs are checked again and stored separately from the sampled terminal graphs; they do not replace the terminal graph or reward used in the trajectory-balance loss. 

\section{Target states}
\label{sec:si_grinch_target_states}

We list the target states using the notation defined below. In the explicit target expansions, an overall normalization factor is omitted: an expansion containing $K$ distinct computational-basis terms with coefficients of unit magnitude is normalized by multiplying it by $1/\sqrt{K}$. All relative signs are retained. Ancilla nodes appear after a tensor product; thus, $\ket{0}$, $\ket{00}$, and $\ket{0000}$ denote one, two, and four ancilla nodes, respectively, each in mode $0$. 

The notation $\ket{GHZ(n,d)}$ denotes a Greenberger--Horne--Zeilinger state of $n$ non-ancilla nodes, each with $d$ local modes labeled $0,\ldots,d-1$. Its normalized form is
\begin{equation*}
\ket{GHZ(n,d)}=\frac{1}{\sqrt{d}}\sum_{j=0}^{d-1}\ket{j}^{\otimes n}.
\end{equation*}
Thus, the first argument specifies the number of nodes, whereas the second specifies the local dimension.

For a Dicke state $\ket{D(n,(n_0,n_1,\ldots))}$, $n_j$ denotes the number of nodes in mode $j$, with $\sum_j n_j=n$. The state is an equal superposition of all distinct computational-basis strings with these occupation numbers. In particular, $\ket{D(4,(3,1))}=\ket{W(4)}$.

The notation $\ket{SRV(r_1,r_2,r_3)}$ labels a three-node pure target with Schmidt-rank vector $(r_1,r_2,r_3)$. For the normalized non-ancilla target $\ket{\psi}$, the entries are defined by
\begin{equation*}
r_i=\operatorname{rank}(\rho_i),\qquad \rho_i=\operatorname{Tr}_{\overline{i}}\!\left(\ket{\psi}\bra{\psi}\right),\qquad i=1,2,3,
\end{equation*}
where $\overline{i}$ denotes the other two non-ancilla nodes. Each $r_i$ is the Schmidt rank across the bipartition separating node $i$ from the remaining two nodes, equivalently the number of nonzero Schmidt coefficients for that bipartition. For the targets listed here, the node ordering gives $r_1\geq r_2\geq r_3$. These ranks characterize the local entanglement dimensions but do not uniquely determine the state; the explicit expansions below specify the particular representatives used. Separable ancilla nodes are not included in the SRV label. We distinguish the four-term cluster-state representation $\ket{C_{\mathrm{1D}}(4)}$ from the sixteen-term path-graph representation $\ket{G_{\mathrm{lin}}(4)}$ to specify the computational basis used in each calculation.  

For photonic gates, the comma in $\ket{\mathrm{in},\mathrm{out}}$ separates the input and output registers. The arguments of each gate abbreviation give the local dimensions of one register. Here, $CNOT(d_c,d_t)$ implements $\ket{a,b}\mapsto\ket{a,(b+a)\bmod d_t}$. The gate $Toffoli(2,2,d_t)$ increments the target modulo $d_t$ only when both controls equal $1$. The two-qubit gate $U_{\mathrm{sq}}(2,2)=CZ(H\otimes H)CZ$, where $H$ is the Hadamard gate, denotes the target used for the square-cluster calculation in quantum-gate mode.

\begin{itemize}
\item $\ket{GHZ(3,2)}\otimes\ket{0}$.
\begin{equation*}
\begin{aligned}
&\bigl(\ket{000} + \ket{111}\bigr)\otimes\ket{0}.
\end{aligned}
\end{equation*}

\item $\ket{GHZ(4,2)}$.
\begin{equation*}
\begin{aligned}
&\ket{0000} + \ket{1111}.
\end{aligned}
\end{equation*}

\item $\ket{GHZ(6,2)}$.
\begin{equation*}
\begin{aligned}
&\ket{000000} + \ket{111111}.
\end{aligned}
\end{equation*}

\item $\ket{GHZ(6,3)^*}$.
\begin{equation*}
\begin{aligned}
&\ket{000000} + \ket{111111} + \ket{222222}.
\end{aligned}
\end{equation*}

\item $\ket{GHZ(8,2)}$.
\begin{equation*}
\begin{aligned}
&\ket{00000000} + \ket{11111111}.
\end{aligned}
\end{equation*}

\item $\ket{D(3,(1,1,1))}\otimes\ket{0}$.
\begin{equation*}
\begin{aligned}
&\bigl(\ket{012} + \ket{021} + \ket{102} + \ket{120} \\
&\quad + \ket{201} + \ket{210}\bigr)\otimes\ket{0}.
\end{aligned}
\end{equation*}

\item $\ket{D(4,(2,2))}$.
\begin{equation*}
\begin{aligned}
&\ket{0011} + \ket{0101} + \ket{0110} + \ket{1001} \\
&\quad + \ket{1010} + \ket{1100}.
\end{aligned}
\end{equation*}

\item $\ket{W(4)}$ (equivalently, $\ket{D(4,(3,1))}$).
\begin{equation*}
\begin{aligned}
&\ket{0001} + \ket{0010} + \ket{0100} + \ket{1000}.
\end{aligned}
\end{equation*}

\item $\ket{SRV(5,4,4)}\otimes\ket{0}$.
\begin{equation*}
\begin{aligned}
&\bigl(\ket{000} + \ket{111} + \ket{222} + \ket{330} \\
&\quad + \ket{413}\bigr)\otimes\ket{0}.
\end{aligned}
\end{equation*}

\item $\ket{SRV(5,4,4)^*}\otimes\ket{0}$.
\begin{equation*}
\begin{aligned}
&\bigl(\ket{000} + \ket{111} + \ket{222} + \ket{333} \\
&\quad + \ket{401}\bigr)\otimes\ket{0}.
\end{aligned}
\end{equation*}

\item $\ket{SRV(5,4,4)^*}\otimes\ket{000}$.
\begin{equation*}
\begin{aligned}
&\bigl(\ket{000} + \ket{111} + \ket{222} + \ket{333} \\
&\quad + \ket{401}\bigr)\otimes\ket{000}.
\end{aligned}
\end{equation*}

\item $\ket{SRV(6,4,4)}\otimes\ket{0}$.
\begin{equation*}
\begin{aligned}
&\bigl(\ket{000} + \ket{111} + \ket{222} + \ket{330} \\
&\quad + \ket{413} + \ket{512}\bigr)\otimes\ket{0}.
\end{aligned}
\end{equation*}

\item $\ket{SRV(6,4,4)^*}\otimes\ket{0}$.
\begin{equation*}
\begin{aligned}
&\bigl(\ket{000} + \ket{111} + \ket{222} + \ket{310} \\
&\quad + \ket{420} + \ket{533}\bigr)\otimes\ket{0}.
\end{aligned}
\end{equation*}

\item $\ket{SRV(6,4,4)^*}\otimes\ket{000}$.
\begin{equation*}
\begin{aligned}
&\bigl(\ket{000} + \ket{111} + \ket{222} + \ket{310} \\
&\quad + \ket{420} + \ket{533}\bigr)\otimes\ket{000}.
\end{aligned}
\end{equation*}

\item $\ket{SRV(6,5,4)}\otimes\ket{0}$.
\begin{equation*}
\begin{aligned}
&\bigl(\ket{000} + \ket{111} + \ket{222} + \ket{330} \\
&\quad + \ket{440} + \ket{513}\bigr)\otimes\ket{0}.
\end{aligned}
\end{equation*}

\item $\ket{SRV(8,6,5)}\otimes\ket{0}$. \small{This state is labeled as SRV(9,5,5) in Ref. \cite{Klaus} but its ordered ranks correspond to 8,6,5.}
\begin{equation*}
\begin{aligned}
&\bigl(\ket{000} + \ket{111} + \ket{222} + \ket{303} \\
&\quad + \ket{404} + \ket{505} + \ket{631} \\
&\quad + \ket{741} + \ket{841}\bigr)\otimes\ket{0}.
\end{aligned}
\end{equation*}

\item $\ket{G(2)}$.
\begin{equation*}
\begin{aligned}
&\ket{00} + \ket{01} + \ket{10} - \ket{11}.
\end{aligned}
\end{equation*}

\item $\ket{G_{\mathrm{lin}}(3)}\otimes\ket{0}$.
\begin{equation*}
\begin{aligned}
&\bigl(\ket{000} + \ket{001} + \ket{010} - \ket{011} \\
&\quad + \ket{100} + \ket{101} - \ket{110} + \ket{111}\bigr)\otimes\ket{0}.
\end{aligned}
\end{equation*}

\item $\ket{C_{\mathrm{1D}}(4)}$.
\begin{equation*}
\begin{aligned}
&\ket{0000} + \ket{0011} + \ket{1100} - \ket{1111}.
\end{aligned}
\end{equation*}

\item $\ket{G_{\mathrm{lin}}(4)}$.
\begin{equation*}
\begin{aligned}
&\ket{0000} + \ket{0001} + \ket{0010} - \ket{0011} \\
&\quad + \ket{0100} + \ket{0101} - \ket{0110} + \ket{0111} \\
&\quad + \ket{1000} + \ket{1001} + \ket{1010} - \ket{1011} \\
&\quad - \ket{1100} - \ket{1101} + \ket{1110} - \ket{1111}.
\end{aligned}
\end{equation*}

\item $\ket{C_{\mathrm{2D}}(4)}$.
\begin{equation*}
\begin{aligned}
&\ket{0000} + \ket{0001} + \ket{0010} - \ket{0011} \\
&\quad + \ket{0100} - \ket{0101} + \ket{0110} + \ket{0111} \\
&\quad + \ket{1000} + \ket{1001} - \ket{1010} + \ket{1011} \\
&\quad - \ket{1100} + \ket{1101} + \ket{1110} + \ket{1111}.
\end{aligned}
\end{equation*}

\item $\ket{G_{\mathrm{star}}(4)}$.
\begin{equation*}
\begin{aligned}
&\ket{0000} + \ket{0001} + \ket{0010} + \ket{0011} \\
&\quad + \ket{0100} + \ket{0101} + \ket{0110} + \ket{0111} \\
&\quad + \ket{1000} - \ket{1001} - \ket{1010} + \ket{1011} \\
&\quad - \ket{1100} + \ket{1101} + \ket{1110} - \ket{1111}.
\end{aligned}
\end{equation*}

\item $\ket{CNOT(2,2)}\otimes\ket{00}$.
\begin{equation*}
\begin{aligned}
&\bigl(\ket{00,00} + \ket{01,01} + \ket{10,11} \\
&\quad + \ket{11,10}\bigr)\otimes\ket{00}.
\end{aligned}
\end{equation*}

\item $\ket{CNOT(2,3)}\otimes\ket{00}$.
\begin{equation*}
\begin{aligned}
&\bigl(\ket{00,00} + \ket{01,01} + \ket{02,02} \\
&\quad + \ket{10,11} + \ket{11,12} + \ket{12,10}\bigr)\otimes\ket{00}.
\end{aligned}
\end{equation*}

\item $\ket{CZ(2,2)}$.
\begin{equation*}
\begin{aligned}
&\ket{00,00} + \ket{01,01} + \ket{10,10} \\
&\quad - \ket{11,11}.
\end{aligned}
\end{equation*}

\item $\ket{CZ(2,2)}\otimes\ket{00}$.
\begin{equation*}
\begin{aligned}
&\bigl(\ket{00,00} + \ket{01,01} + \ket{10,10} \\
&\quad - \ket{11,11}\bigr)\otimes\ket{00}.
\end{aligned}
\end{equation*}

\item $\ket{U_{\mathrm{sq}}(2,2)}\otimes\ket{00}$.
\begin{equation*}
\begin{aligned}
&\bigl(\ket{00,00} + \ket{00,01} + \ket{00,10} \\
&\quad - \ket{00,11} + \ket{01,00} - \ket{01,01} \\
&\quad + \ket{01,10} + \ket{01,11} + \ket{10,00} \\
&\quad + \ket{10,01} - \ket{10,10} + \ket{10,11} \\
&\quad - \ket{11,00} + \ket{11,01} + \ket{11,10} \\
&\quad + \ket{11,11}\bigr)\otimes\ket{00}.
\end{aligned}
\end{equation*}

\item $\ket{Toffoli(2,2,2)}\otimes\ket{00}$.
\begin{equation*}
\begin{aligned}
&\bigl(\ket{000,000} + \ket{001,001} + \ket{010,010} \\
&\quad + \ket{011,011} + \ket{100,100} + \ket{101,101} \\
&\quad + \ket{110,111} + \ket{111,110}\bigr)\otimes\ket{00}.
\end{aligned}
\end{equation*}

\item $\ket{Toffoli(2,2,2)}\otimes\ket{0000}$.
\begin{equation*}
\begin{aligned}
&\bigl(\ket{000,000} + \ket{001,001} + \ket{010,010} \\
&\quad + \ket{011,011} + \ket{100,100} + \ket{101,101} \\
&\quad + \ket{110,111} + \ket{111,110}\bigr)\otimes\ket{0000}.
\end{aligned}
\end{equation*}

\end{itemize}

\paragraph{Local equivalence of the four-qubit cluster and linear graph targets.}
The targets $\ket{C_{\mathrm{1D}}(4)}$ and $\ket{G_{\mathrm{lin}}(4)}$ are related by local Hadamard transformations on the two end qubits \cite{LinearClusterEquiv}. We label the qubits from left to right as $1,2,3,4$ and explicitly include the normalization factors. The four-term cluster target is
\begin{equation}
\ket{C_{\mathrm{1D}}(4)}=\frac{1}{2}\left(\ket{0000}+\ket{0011}+\ket{1100}-\ket{1111}\right).
\label{eq:cluster_four_term_normalized}
\end{equation}
Let $I$ denote the single-qubit identity operator and define
\begin{equation}
U_{\mathrm{end}}=H\otimes I\otimes I\otimes H,
\end{equation}
where the Hadamard gate acts as
\begin{equation}
H\ket{0}=\ket{+}=\frac{\ket{0}+\ket{1}}{\sqrt{2}},\qquad H\ket{1}=\ket{-}=\frac{\ket{0}-\ket{1}}{\sqrt{2}}.
\end{equation}
Applying $U_{\mathrm{end}}$ to the four-term target gives
\begin{equation}
U_{\mathrm{end}}\ket{C_{\mathrm{1D}}(4)}=\frac{1}{2}\left(\ket{+00+}+\ket{+01-}+\ket{-10+}-\ket{-11-}\right).
\label{eq:cluster_end_hadamards}
\end{equation}
Expanding the first and fourth qubits in the computational basis yields
\begin{equation}
\begin{aligned}
U_{\mathrm{end}}\ket{C_{\mathrm{1D}}(4)}
={}&\frac{1}{4}\bigl(
\ket{0000}+\ket{0001}+\ket{0010}-\ket{0011}\\
&\quad+\ket{0100}+\ket{0101}-\ket{0110}+\ket{0111}\\
&\quad+\ket{1000}+\ket{1001}+\ket{1010}-\ket{1011}\\
&\quad-\ket{1100}-\ket{1101}+\ket{1110}-\ket{1111}
\bigr)\\
={}&\ket{G_{\mathrm{lin}}(4)}.
\end{aligned}
\label{eq:cluster_linear_explicit_transformation}
\end{equation}
Since $H^2=I$, the transformation is its own inverse:
\begin{equation}
\ket{C_{\mathrm{1D}}(4)}=U_{\mathrm{end}}\ket{G_{\mathrm{lin}}(4)}.
\label{eq:cluster_linear_inverse_transformation}
\end{equation}
Thus, the two targets are locally Clifford-equivalent, although their computational-basis expansions differ. Their distinct optical-graph edge counts characterize the solutions found for these basis-specific targets, rather than different entanglement classes or minimal resource requirements.

\section{Additional results} \label{sec:additional_results}
\subsection{Strict vs relaxed edge budget}

\begin{figure}[h]
    \centering
    \includegraphics[width=0.99\linewidth]{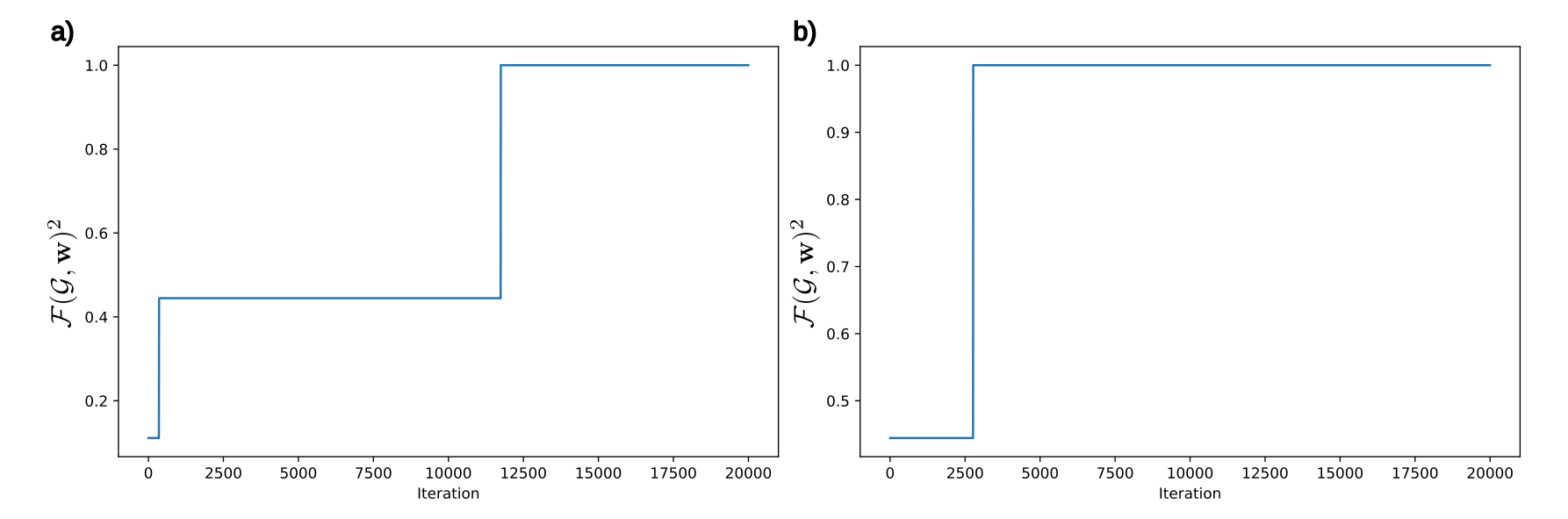}
    \caption[Rewards progress for strict and relaxed edge budgets.]{Rewards progress for the $\ket{SRV(5,4,4)}\otimes\ket{0}$ state with a strict edge budget of 10 edges (panel a) and a relaxed edge budget of 12 edges (panel b).}
    \label{fig:rewards_progress_10vs12}
\end{figure}

Figure~\ref{fig:rewards_progress_10vs12} compares the reward evolution for $\ket{SRV(5,4,4)}\otimes\ket{0}$ using prescribed edge budgets of ten and twelve edges, referred to here as strict and relaxed, respectively. We isolate the effect of trajectory length by removing the logic-based construction masks used in \texttt{Grinch}; thus, the sampling evaluates the effect of the trajectory on our algorithm's learning capabilities. The relaxed run reaches high-reward solutions earlier in this comparison. This observation is consistent with additional construction flexibility facilitating the search.

\subsection{Comparison with other approaches}
In Table \ref{tab:State-Fidelities-vs}, we compare the solutions found with \texttt{Grinch} against those of similar solvers. We highlight the lower variance in the number of edges obtained with \texttt{Grinch}; for the selected states, we consistently find the same number of edges for the top sampled state with GFlowNets across independent runs.

\begin{table}[h!]
    \centering
    \scriptsize
    \renewcommand{\arraystretch}{1.15}
    \setlength{\tabcolsep}{2.0pt}
    \begin{tabular}{@{}lccccc@{}}
        & \multicolumn{2}{c}{Grinch}
        & \multicolumn{3}{c}{Ref.~\cite{Klaus}} \\
        \cmidrule(lr){2-3}\cmidrule(l){4-6}
        \textbf{Target state} & $\mathcal{F}$ & $|E|$
                     & $\overline{\mathcal{F}}$
                     & $\overline{|E|}\!\pm\!\sigma_E$
                     & \textbf{Method} \\ \hline
        $\ket{GHZ(6,2)}$                      & 1.00 & 6  & 1.00 & $6.0\!\pm\!0.0$   & K/KO \\
        $\ket{GHZ(6,3)^*}$                    & 1.00 & 9  & 1.00 & $9.2\!\pm\!0.0$   & TO \\
        \hline
        $\ket{SRV(5,4,4)}\otimes\ket{0}$     & 1.00 & 8  & 1.00 & $8.0\!\pm\!0.0$   & K/TO/KO \\
        $\ket{SRV(5,4,4)^*}\otimes\ket{0}$   & 0.80 & 18 & 0.80 & $17.6\!\pm\!12.4$ & KO \\
        $\ket{SRV(6,4,4)}\otimes\ket{0}$     & 1.00 & 9 & 1.00 & $8.0\!\pm\!0.0$   & TO \\
        $\ket{SRV(6,4,4)^*}\otimes\ket{0}$   & 0.83 & 18 & 0.83 & $20.3\!\pm\!14.3$ & TO \\
        $\ket{SRV(6,5,4)}\otimes\ket{0}$     & 1.00 & 9 & 1.00 & $9.0\!\pm\!0.0$   & TO/KO \\
    \end{tabular}
    \caption{Performance comparison of \texttt{Grinch} relaxed with Ref.~\cite{Klaus} reported methods; Klaus (K), Theseus (T), TheseusOpt (TO), or KlausOpt (KO). We select the method with the largest mean fidelity and, among fidelity ties, the smallest mean edge count.}
    \label{tab:State-Fidelities-vs}
\end{table}

\subsection{Analytic SRV constructions from topologies discovered by \texttt{Grinch}}
\label{sec:srv-constructions}

\subsubsection{Targets and normalization}

We construct analytic weight families on the 14- and 15-edge colored topologies labeled $\max\mathcal{F}$ in Fig.~\ref{fig:SRVs3a}. Denote these graphs by $\mathcal{G}_{\alpha}$, with $\alpha\in\{544^*,644^*\}$, and define the normalized targets
\begin{align}
\ket{T_{544^*}}&\equiv\ket{SRV(5,4,4)^*}\otimes\ket{000},
\label{eq:srv-target544}\\
\ket{T_{644^*}}&\equiv\ket{SRV(6,4,4)^*}\otimes\ket{000}.
\label{eq:srv-target644}
\end{align}
Their basis expansions are given in Sec.~\ref{sec:si_grinch_target_states}, with normalization factors $1/\sqrt{5}$ and $1/\sqrt{6}$, respectively. The digit order is the vertex order $(0,1,2,3,4,5)$: vertices $0,1,2$ carry the target and vertices $3,4,5$ are ancillas in mode $0$. The analytic assignments below use the edge sets of the displayed numerical solutions; the diamonds in the figure indicate the signs of the numerical weights.

We use the edge convention $e_i=((u_i,v_i),(c_{u_i},c_{v_i}))$ and write $w_i\equiv w_{u_i,v_i}^{c_{u_i},c_{v_i}}$, with edge indices local to each topology. Using the matching amplitudes $A_b(\mathcal{G},\vw)$ defined in Eq.~\eqref{eq:pruning_matching_amplitude}, the unnormalized postselected state along each family is
\begin{equation}
\ket{\widetilde\psi_{\alpha}(t)}
=\sum_b A_b\bigl(\mathcal{G}_{\alpha},\vw_{\alpha}(t)\bigr)\ket{b}.
\label{eq:srv-pm-state}
\end{equation}
For a nonzero output, define
\begin{equation}
\begin{aligned}
Q_{\alpha}(t)&=\langle\widetilde\psi_{\alpha}(t)|\widetilde\psi_{\alpha}(t)\rangle,\\
\ket{\psi_{\alpha}(t)}&=\frac{\ket{\widetilde\psi_{\alpha}(t)}}{\sqrt{Q_{\alpha}(t)}},\\
\mathcal{F}_{\alpha}(t)&=\frac{|\langle T_{\alpha}|\widetilde\psi_{\alpha}(t)\rangle|^2}{Q_{\alpha}(t)}.
\end{aligned}
\label{eq:srv-normalization}
\end{equation}
We take $0<t\leq1$, so all edge weights remain nonzero in both assignments, with $\max_i|w_i(t)|=1$. Each assignment has three negative weights, corresponding to relative phases of $\pi$. The tables below show how the 13 perfect matchings of each graph contribute to the ten computational-basis states $\ket{b}$ and combine to give their amplitudes. We leave the graph dependence of $A_b$ implicit because each table refers to a fixed topology.

\subsubsection{The 14-edge $\ket{SRV(5,4,4)^*}$ construction}
\label{sec:srv544-family}

Table~\ref{tab:srv544-edges} defines $\vw_{544^*}(t)$, and Table~\ref{tab:srv544-amplitudes} gives the resulting amplitudes.

\begin{table}[htbp]
\centering
\scriptsize
\caption{Colored edges and analytic weights of $\mathcal{G}_{544^*}$. Endpoint modes are ordered as $(u_i,v_i)$.}
\label{tab:srv544-edges}
\begin{tabular}{rccc}
\toprule
$i$ & $(u_i,v_i)$ & $(c_{u_i},c_{v_i})$ & $w_i(t)$\\
\midrule
1 &(0,1)&(1,1)&$t$\\
2 &(0,2)&(0,0)&$t$\\
3 &(0,2)&(3,3)&$-1$\\
4 &(0,2)&(4,1)&$t$\\
5 &(0,3)&(3,0)&$1$\\
6 &(0,4)&(2,0)&$1$\\
7 &(1,2)&(2,2)&$t$\\
8 &(1,2)&(3,3)&$-1$\\
9 &(1,3)&(3,0)&$1$\\
10 &(1,4)&(0,0)&$1$\\
11 &(2,4)&(1,0)&$1$\\
12 &(2,5)&(3,0)&$1$\\
13 &(3,5)&(0,0)&$1$\\
14 &(4,5)&(0,0)&$-t/2$\\
\bottomrule
\end{tabular}
\end{table}

\begin{table}[htbp]
\centering
\scriptsize
\caption{Matching amplitudes for $\mathcal{G}_{544^*}$ before and after the weight assignment in Table~\ref{tab:srv544-edges}.}
\label{tab:srv544-amplitudes}
\begin{tabular}{clc}
\toprule
$\ket{b}$ & $A_b(\vw)$ & $A_b(\vw_{544^*}(t))$\\
\midrule
$\ket{000000}$ & $w_2w_{10}w_{13}$ & $t$\\
$\ket{030000}$ & $w_2w_9w_{14}$ & $-t^2/2$\\
$\ket{111000}$ & $w_1w_{11}w_{13}$ & $t$\\
$\ket{222000}$ & $w_6w_7w_{13}$ & $t$\\
$\ket{233000}$ & $w_6w_8w_{13}+w_6w_9w_{12}$ & $0$\\
$\ket{303000}$ & $w_3w_{10}w_{13}+w_5w_{10}w_{12}$ & $0$\\
$\ket{322000}$ & $w_5w_7w_{14}$ & $-t^2/2$\\
$\ket{333000}$ & $w_3w_9w_{14}+w_5w_8w_{14}$ & $t$\\
$\ket{401000}$ & $w_4w_{10}w_{13}$ & $t$\\
$\ket{431000}$ & $w_4w_9w_{14}$ & $-t^2/2$\\
\bottomrule
\end{tabular}
\end{table}

The relations $w_8w_{13}+w_9w_{12}=0$ and $w_3w_{13}+w_5w_{12}=0$ cancel the intruder components $\ket{233000}$ and $\ket{303000}$, respectively. The two matching contributions to $\ket{333000}$ add to $t$. All five target components therefore have amplitude $t$, while the three surviving intruders have amplitude $-t^2/2$:
\begin{equation}
\begin{aligned}
\ket{\widetilde\psi_{544^*}(t)}={}&\sqrt{5}\,t\ket{T_{544^*}}\\
&-\frac{t^2}{2}\bigl(\ket{030000}+\ket{322000}+\ket{431000}\bigr).
\end{aligned}
\label{eq:srv544-raw}
\end{equation}
Orthogonality of the target and intruder components gives
\begin{equation}
\begin{aligned}
Q_{544^*}(t)&=5t^2+\frac{3t^4}{4},\\
\mathcal{F}_{544^*}(t)&=\frac{5t^2}{Q_{544^*}(t)}=\frac{20}{20+3t^2}.
\end{aligned}
\label{eq:srv544-fidelity}
\end{equation}

\subsubsection{The 15-edge $\ket{SRV(6,4,4)^*}$ construction}
\label{sec:srv644-family}

Table~\ref{tab:srv644-edges} defines $\vw_{644^*}(t)$, with matching amplitudes listed in Table~\ref{tab:srv644-amplitudes}.

\begin{table}[htbp]
\centering
\scriptsize
\caption{Colored edges and analytic weights of $\mathcal{G}_{644^*}$. Endpoint modes are ordered as $(u_i,v_i)$.}
\label{tab:srv644-edges}
\begin{tabular}{rccc}
\toprule
$i$ & $(u_i,v_i)$ & $(c_{u_i},c_{v_i})$ & $w_i(t)$\\
\midrule
1 &(0,1)&(0,0)&$t$\\
2 &(0,1)&(4,2)&$t$\\
3 &(0,2)&(1,1)&$t$\\
4 &(0,2)&(5,3)&$-1$\\
5 &(0,3)&(2,0)&$t$\\
6 &(0,3)&(5,0)&$-1$\\
7 &(0,4)&(3,0)&$t$\\
8 &(0,5)&(5,0)&$1$\\
9 &(1,2)&(2,2)&$1$\\
10 &(1,3)&(1,0)&$1$\\
11 &(1,5)&(3,0)&$-t/2$\\
12 &(2,4)&(3,0)&$1$\\
13 &(2,5)&(0,0)&$1$\\
14 &(3,4)&(0,0)&$1$\\
15 &(4,5)&(0,0)&$1$\\
\bottomrule
\end{tabular}
\end{table}

\begin{table}[htbp]
\centering
\scriptsize
\caption{Matching amplitudes for $\mathcal{G}_{644^*}$ before and after the weight assignment in Table~\ref{tab:srv644-edges}.}
\label{tab:srv644-amplitudes}
\begin{tabular}{clc}
\toprule
$\ket{b}$ & $A_b(\vw)$ & $A_b(\vw_{644^*}(t))$\\
\midrule
$\ket{000000}$ & $w_1w_{13}w_{14}$ & $t$\\
$\ket{111000}$ & $w_3w_{10}w_{15}$ & $t$\\
$\ket{131000}$ & $w_3w_{11}w_{14}$ & $-t^2/2$\\
$\ket{222000}$ & $w_5w_9w_{15}$ & $t$\\
$\ket{233000}$ & $w_5w_{11}w_{12}$ & $-t^2/2$\\
$\ket{310000}$ & $w_7w_{10}w_{13}$ & $t$\\
$\ket{420000}$ & $w_2w_{13}w_{14}$ & $t$\\
$\ket{513000}$ & $w_4w_{10}w_{15}+w_8w_{10}w_{12}$ & $0$\\
$\ket{522000}$ & $w_6w_9w_{15}+w_8w_9w_{14}$ & $0$\\
$\ket{533000}$ & $w_4w_{11}w_{14}+w_6w_{11}w_{12}$ & $t$\\
\bottomrule
\end{tabular}
\end{table}

Here $w_4w_{15}+w_8w_{12}=0$ and $w_6w_{15}+w_8w_{14}=0$ cancel the intruders $\ket{513000}$ and $\ket{522000}$. The two contributions to $\ket{533000}$ add to $t$, giving six target amplitudes $t$ and two surviving intruder amplitudes $-t^2/2$:
\begin{equation}
\begin{aligned}
\ket{\widetilde\psi_{644^*}(t)}={}&\sqrt{6}\,t\ket{T_{644^*}}\\
&-\frac{t^2}{2}\bigl(\ket{131000}+\ket{233000}\bigr),
\end{aligned}
\label{eq:srv644-raw}
\end{equation}
and hence
\begin{equation}
\begin{aligned}
Q_{644^*}(t)&=6t^2+\frac{t^4}{2},\\
\mathcal{F}_{644^*}(t)&=\frac{6t^2}{Q_{644^*}(t)}=\frac{12}{12+t^2}.
\end{aligned}
\label{eq:srv644-fidelity}
\end{equation}

Both normalized states approach their targets as $t\to0^+$, while $Q_{\alpha}(t)\to0$. At $t=0$, the unnormalized states vanish and conditional normalization is undefined. Neither topology admits a nonzero exact target output, even with arbitrary complex edge weights. For $\mathcal{G}_{544^*}$, the required nonzero amplitudes $A_{000000}$, $A_{222000}$, and $A_{333000}$ imply $w_2,w_7,w_{14}\neq0$. Eliminating $A_{030000}=w_2w_9w_{14}$ and $A_{322000}=w_5w_7w_{14}$ then forces $w_9=w_5=0$, making the required amplitude $A_{333000}=w_{14}(w_3w_9+w_5w_8)$ vanish. For $\mathcal{G}_{644^*}$, nonzero $A_{111000}$ and $A_{000000}$ require $w_3,w_{14}\neq0$. Eliminating $A_{131000}=w_3w_{11}w_{14}$ forces $w_{11}=0$, which also eliminates the required amplitude $A_{533000}=w_{11}(w_4w_{14}+w_6w_{12})$.

\subsubsection{Fidelity and sixfold coincidence rates}
\label{sec:srv-rates}

For each family $\alpha\in\{544^*,644^*\}$, we write the physical pair amplitudes as $g_{\alpha}\vw_{\alpha}(t)$. Since $\max_i|w_i(t)|=1$, the common scale $|g_{\alpha}|\ll1$ is the largest pair-amplitude magnitude. In the weak-pair-generation model~\cite{PyTheus}, each perfect matching involves the creation of three photon pairs. The leading-order probability per pump pulse of producing one photon in each of the six output paths is therefore
\begin{equation}
P_{\mathrm{post},\alpha}(t)\simeq|g_{\alpha}|^6Q_{\alpha}(t).
\label{eq:srv-physical-scale}
\end{equation}
Let $f_{\mathrm{rep},\alpha}$ be the pump repetition rate and $\eta_{\alpha,j}$ the combined transmission and detection efficiency for path $j$, assumed independent of the photon's internal mode. The detected sixfold coincidence rate is then
\begin{equation}
\begin{aligned}
R_{\alpha}(t)&\simeq\kappa_{\alpha}Q_{\alpha}(t),\\
\kappa_{\alpha}&=f_{\mathrm{rep},\alpha}|g_{\alpha}|^6\prod_{j=0}^{5}\eta_{\alpha,j}.
\end{aligned}
\label{eq:srv-rate-prefactor}
\end{equation}

We first keep the source scale, repetition rate, and path efficiencies fixed. Eliminating $t$ from Eqs.~\eqref{eq:srv544-fidelity} and~\eqref{eq:srv644-fidelity} expresses the rates directly in terms of the conditional fidelities:
\begin{align}
R_{544^*}(t)&\simeq\frac{100}{3}\kappa_{544^*}\frac{1-\mathcal{F}_{544^*}(t)}{\mathcal{F}_{544^*}(t)^2},
\label{eq:srv544-rate}\\
R_{644^*}(t)&\simeq72\kappa_{644^*}\frac{1-\mathcal{F}_{644^*}(t)}{\mathcal{F}_{644^*}(t)^2}.
\label{eq:srv644-rate}
\end{align}
Thus, the rates vanish linearly as the infidelity $\delta_{\alpha}(t)\equiv1-\mathcal{F}_{\alpha}(t)$ tends to zero. For example, reducing the infidelity from $10^{-5}$ to $10^{-6}$ lowers the predicted rate by approximately a factor of ten along either family.

We can also hold the total pair-generation strength fixed by allowing the common source scale to vary with $t$. In the same leading-order model, this strength is $p_{\alpha}=|g_{\alpha}(t)|^2S_{\alpha}(t)\ll1$, where $S_{\alpha}(t)=\|\vw_{\alpha}(t)\|_2^2$. The analytic weight assignments give
\begin{equation}
\begin{aligned}
S_{544^*}(t)&=9+\frac{17}{4}t^2,\\
S_{644^*}(t)&=9+\frac{21}{4}t^2.
\end{aligned}
\label{eq:srv-total-strength}
\end{equation}
Keeping $p_{\alpha}$ fixed requires $|g_{\alpha}(t)|^2=p_{\alpha}/S_{\alpha}(t)$. Substituting this relation into Eq.~\eqref{eq:srv-rate-prefactor} gives
\begin{equation}
R_{\alpha}(t)\simeq f_{\mathrm{rep},\alpha}\left(\prod_{j=0}^{5}\eta_{\alpha,j}\right)p_{\alpha}^3\frac{Q_{\alpha}(t)}{S_{\alpha}(t)^3}.
\label{eq:srv-fixed-total-rate}
\end{equation}
Since $S_{\alpha}(t)\to9$ as $t\to0^+$, the rates remain asymptotically linear in infidelity when the repetition rate and path efficiencies are fixed.

The rates above count all postselected sixfold events. If the detection measurement also implements an ideal projection onto $\ket{T_{\alpha}}$, the corresponding outcome rate is $R_{T,\alpha}(t)=\mathcal{F}_{\alpha}(t)R_{\alpha}(t)$. Predicting an absolute count rate requires calibrated source and detection parameters, together with the pump repetition rate. These expressions apply while higher-pair emissions and accidental coincidences remain negligible.

\subsection{Sampling rates} \label{sec:sampling-rates}

\begin{figure}[h!]
    \centering
    \includegraphics[width=0.8\linewidth]{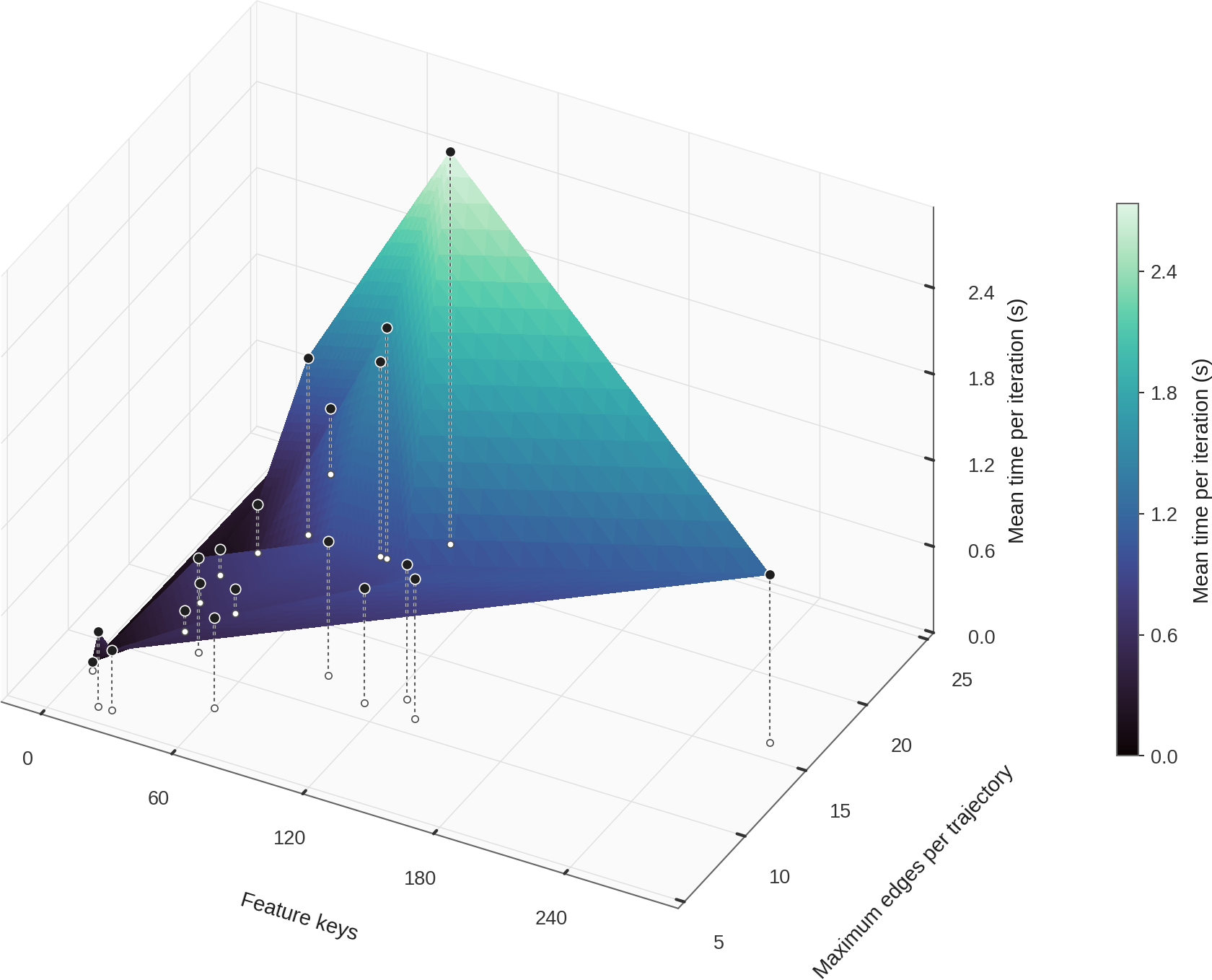}
    \caption{\texttt{Grinch} training time across feature counts and trajectory lengths. Filled markers show the mean training time per iteration, averaged across runs that share the same feature count and maximum edge combination. The surface provides piecewise-linear interpolation within the sampled region. Dashed guides connect observations to their projections onto the plane (open circles) and are overlaid for visibility.}
    \label{fig:timing}
\end{figure}

Here, we show in Fig.~\ref{fig:timing} the time per iteration for \texttt{Grinch} as a function of the maximum number of edges in the trajectory and the number of feature keys. As shown in Section \ref{sec:training_and_architecture}, the number of parameters for each model depends on $|\mathcal{A}|=\operatorname{len}(\texttt{FEATURE\_KEYS})$, thus affecting the time for model optimization.

 \newpage
\bibliography{references}

@article{UsDiscreteFlowbased,
    author = {Huidobro-Meezs, Isaac L. and Dai, Jun and Vargas-Hernández, Rodrigo A.},
    title = {Discrete flow-based generative models for measurement optimization in quantum computing},
    journal = {Digital Discovery},
    volume = {5},
    number = {11},
    pages = {5078-5096},
    year = {2026},
    month = {11},
    issn = {2635-098X},
    doi = {10.1039/d6dd00430j},
    url = {https://doi.org/10.1039/d6dd00430j},
    eprint = {https://pubs.rsc.org/dd/article-pdf/5/11/5078/14865258/d6dd00430j.pdf},
}

@article{bengio2021flow,
  title={Flow network based generative models for non-iterative diverse candidate generation},
  author={Bengio, Emmanuel and Jain, Moksh and Korablyov, Maksym and Precup, Doina and Bengio, Yoshua},
  journal={Advances in Neural Information Processing Systems},
  volume={34},
  pages={27381--27394},
  year={2021}
}

@article{bengio2023gflownet,
  title={G{F}low{N}et foundations},
  author={Bengio, Yoshua and Lahlou, Salem and Deleu, Tristan and Hu, Edward J and Tiwari, Mo and Bengio, Emmanuel},
  journal={The Journal of Machine Learning Research},
  volume={24},
  number={1},
  pages={10006--10060},
  year={2023},
  publisher={JMLRORG}
}

@article{jain2023gflownets,
  title={G{F}low{N}ets for AI-driven scientific discovery},
  author={Jain, Moksh and Deleu, Tristan and Hartford, Jason and Liu, Cheng-Hao and Hernandez-Garcia, Alex and Bengio, Yoshua},
  journal={Digital Discovery},
  volume={2},
  number={3},
  pages={557--577},
  year={2023},
  publisher={Royal Society of Chemistry}
}

@article{zhang2023letFlows,
  title={Let the flows tell: Solving graph combinatorial problems with {GF}low{N}ets},
  author={Zhang, Dinghuai and Dai, Hanjun and Malkin, Nikolay and Courville, Aaron C and Bengio, Yoshua and Pan, Ling},
  journal={Advances in neural information processing systems},
  volume={36},
  pages={11952--11969},
  year={2023}
}

@article{malkin2022trajectory,
  title={Trajectory balance: Improved credit assignment in {GF}low{N}ets},
  author={Malkin, Nikolay and Jain, Moksh and Bengio, Emmanuel and Sun, Chen and Bengio, Yoshua},
  journal={Advances in Neural Information Processing Systems},
  volume={35},
  pages={5955--5967},
  year={2022}
}

@InProceedings{bioseq,
  title = 	 {Biological Sequence Design with {GF}low{N}ets},
  author =       {Jain, Moksh and Bengio, Emmanuel and Hernandez-Garcia, Alex and Rector-Brooks, Jarrid and Dossou, Bonaventure F. P. and Ekbote, Chanakya Ajit and Fu, Jie and Zhang, Tianyu and Kilgour, Michael and Zhang, Dinghuai and Simine, Lena and Das, Payel and Bengio, Yoshua},
  booktitle = 	 {Proceedings of the 39th International Conference on Machine Learning},
  pages = 	 {9786--9801},
  year = 	 {2022},
  editor = 	 {Chaudhuri, Kamalika and Jegelka, Stefanie and Song, Le and Szepesvari, Csaba and Niu, Gang and Sabato, Sivan},
  volume = 	 {162},
  series = 	 {Proceedings of Machine Learning Research},
  month = 	 {17--23 Jul},
  publisher =    {PMLR},
  url = 	 {https://proceedings.mlr.press/v162/jain22a.html}
}

@article{GenerativeQuantumStateReconstruction,
  title = {Reconstructing quantum states with generative models},
  author = {Carrasquilla, Juan and Torlai, Giacomo and Melko, Roger G. and Aolita, Leandro},
  journal = {Nature Machine Intelligence},
  volume = {1},
  pages = {155--161},
  year = {2019},
  doi = {10.1038/s42256-019-0028-1},
  url = {https://doi.org/10.1038/s42256-019-0028-1}
}

@article{LSTMQuantumOptics,
  title = {Quantum Optical Experiments Modeled by Long Short-Term Memory},
  author = {Adler, Thomas and Erhard, Manuel and Krenn, Mario and Brandstetter, Johannes and Kofler, Johannes and Hochreiter, Sepp},
  journal = {Photonics},
  volume = {8},
  number = {12},
  pages = {535},
  year = {2021},
  doi = {10.3390/photonics8120535},
  url = {https://doi.org/10.3390/photonics8120535}
}

@inproceedings{SequentialGraphQuantumOptics,
  author = {{Anonymous}},
  title = {Automatic Quantum Optics Experimental Design with Sequential Graph Generative Models},
  booktitle = {AAAI Workshop on Deep Learning on Graphs: Methodologies and Applications},
  year = {2020},
  note = {Workshop paper},
  url = {https://deep-learning-graphs.bitbucket.io/dlg-aaai20/accepted_papers/DLGMA_2020_paper_36.pdf}
}

@article{QOVAE,
  title = {Learning interpretable representations of entanglement in quantum optics experiments using deep generative models},
  author = {Flam-Shepherd, Daniel and Wu, Tony C. and Gu, Xuemei and Cervera-Lierta, Alba and Krenn, Mario and Aspuru-Guzik, Al{\'{a}}n},
  journal = {Nature Machine Intelligence},
  volume = {4},
  pages = {544--554},
  year = {2022},
  doi = {10.1038/s42256-022-00493-5},
  url = {https://doi.org/10.1038/s42256-022-00493-5}
}

@article{TheseusDesign,
  title = {Conceptual Understanding through Efficient Automated Design of Quantum Optical Experiments},
  author = {Krenn, Mario and Kottmann, Jakob S. and Tischler, Nora and Aspuru-Guzik, Al\'an},
  journal = {Phys. Rev. X},
  volume = {11},
  issue = {3},
  pages = {031044},
  numpages = {15},
  year = {2021},
  month = {Aug},
  publisher = {American Physical Society},
  doi = {10.1103/PhysRevX.11.031044},
  url = {https://link.aps.org/doi/10.1103/PhysRevX.11.031044}
}

@article{QuantumExpAndGraphs,
  title = {Quantum Experiments and Graphs: Multiparty States as Coherent Superpositions of Perfect Matchings},
  author = {Krenn, Mario and Gu, Xuemei and Zeilinger, Anton},
  journal = {Phys. Rev. Lett.},
  volume = {119},
  issue = {24},
  pages = {240403},
  numpages = {6},
  year = {2017},
  month = {Dec},
  publisher = {American Physical Society},
  doi = {10.1103/PhysRevLett.119.240403},
  url = {https://link.aps.org/doi/10.1103/PhysRevLett.119.240403}
}

@article{PossibleStates,
  title = {Quantum experiments and graphs. III. High-dimensional and multiparticle entanglement},
  author = {Gu, Xuemei and Chen, Lijun and Zeilinger, Anton and Krenn, Mario},
  journal = {Phys. Rev. A},
  volume = {99},
  issue = {3},
  pages = {032338},
  numpages = {11},
  year = {2019},
  month = {Mar},
  publisher = {American Physical Society},
  doi = {10.1103/PhysRevA.99.032338},
  url = {https://link.aps.org/doi/10.1103/PhysRevA.99.032338}
}

@article{Melvin,
  title = {Automated Search for new Quantum Experiments},
  author = {Krenn, Mario and Malik, Mehul and Fickler, Robert and Lapkiewicz, Radek and Zeilinger, Anton},
  journal = {Phys. Rev. Lett.},
  volume = {116},
  issue = {9},
  pages = {090405},
  numpages = {5},
  year = {2016},
  month = {Mar},
  publisher = {American Physical Society},
  doi = {10.1103/PhysRevLett.116.090405},
  url = {https://link.aps.org/doi/10.1103/PhysRevLett.116.090405}
}

@article{KnottSearch,
  author = {Knott, Paul A.},
  title = {A search algorithm for quantum state engineering and metrology},
  journal = {New Journal of Physics},
  volume = {18},
  pages = {073033},
  year = {2016}
}

@article{NicholsGenetic,
  author = {Nichols, Rosanna and Mineh, Lama and Rubio, Jes\'{u}s and Matthews, Jonathan C. F. and Knott, Paul A.},
  title = {Designing quantum experiments with a genetic algorithm},
  journal = {Quantum Science and Technology},
  volume = {4},
  pages = {045012},
  year = {2019}
}

@article{ArrazolaPhotonicML,
  author = {Arrazola, Juan Miguel and Bromley, Thomas R. and Izaac, Josh and Myers, Casey R. and Br\'{a}dler, Kamil and Killoran, Nathan},
  title = {Machine learning method for state preparation and gate synthesis on photonic quantum computers},
  journal = {Quantum Science and Technology},
  volume = {4},
  pages = {024004},
  year = {2019}
}

@article{
RL-CircuitSearch,
author = {Alexey A. Melnikov  and Hendrik Poulsen Nautrup  and Mario Krenn  and Vedran Dunjko  and Markus Tiersch  and Anton Zeilinger  and Hans J. Briegel },
title = {Active learning machine learns to create new quantum experiments},
journal = {Proceedings of the National Academy of Sciences},
volume = {115},
number = {6},
pages = {1221-1226},
year = {2018},
doi = {10.1073/pnas.1714936115},
URL = {https://www.pnas.org/doi/abs/10.1073/pnas.1714936115},
eprint = {https://www.pnas.org/doi/pdf/10.1073/pnas.1714936115}}

@article{AI-entanglement,
  title = {Entangling Independent Particles by Path Identity},
  author = {Wang, Kai and Hou, Zhaohua and Qian, Kaiyi and Chen, Leizhen and Krenn, Mario and Zhu, Shining and Ma, Xiao-Song},
  journal = {Phys. Rev. Lett.},
  volume = {133},
  issue = {23},
  pages = {233601},
  numpages = {6},
  year = {2024},
  month = {Dec},
  publisher = {American Physical Society},
  doi = {10.1103/PhysRevLett.133.233601},
  url = {https://link.aps.org/doi/10.1103/PhysRevLett.133.233601}
}

@article{PyTheus,
  doi = {10.22331/q-2023-12-12-1204},
  url = {https://doi.org/10.22331/q-2023-12-12-1204},
  title = {Digital {D}iscovery of 100 diverse {Q}uantum {E}xperiments with {P}y{T}heus},
  author = {Ruiz-Gonzalez, Carlos and Arlt, S{\"{o}}ren and Petermann, Jan and Sayyad, Sharareh and Jaouni, Tareq and Karimi, Ebrahim and Tischler, Nora and Gu, Xuemei and Krenn, Mario},
  journal = {{Quantum}},
  issn = {2521-327X},
  publisher = {{Verein zur F{\"{o}}rderung des Open Access Publizierens in den Quantenwissenschaften}},
  volume = {7},
  pages = {1204},
  month = dec,
  year = {2023}
}

@article{Klaus,
  doi = {10.22331/q-2022-10-13-836},
  url = {https://doi.org/10.22331/q-2022-10-13-836},
  title = {Design of quantum optical experiments with logic artificial intelligence},
  author = {Cervera-Lierta, Alba and Krenn, Mario and Aspuru-Guzik, Al{\'{a}}n},
  journal = {{Quantum}},
  issn = {2521-327X},
  publisher = {{Verein zur F{\"{o}}rderung des Open Access Publizierens in den Quantenwissenschaften}},
  volume = {6},
  pages = {836},
  month = oct,
  year = {2022}
}

@misc{adam,
      title={Adam: A Method for Stochastic Optimization}, 
      author={Diederik P. Kingma and Jimmy Ba},
      year={2017},
      eprint={1412.6980},
      archivePrefix={arXiv},
      primaryClass={cs.LG},
      url={https://arxiv.org/abs/1412.6980}, 
}

@article{LBFGSB,
  author  = {Byrd, Richard H. and Lu, Peihuang and Nocedal, Jorge and Zhu, Ciyou},
  title   = {A Limited Memory Algorithm for Bound Constrained Optimization},
  journal = {SIAM Journal on Scientific Computing},
  volume  = {16},
  number  = {5},
  pages   = {1190--1208},
  year    = {1995},
  doi     = {10.1137/0916069},
  url     = {https://doi.org/10.1137/0916069}
}

@article{InverseDesignPhotonicSystems,
author = {MacLellan, Benjamin and Roztocki, Piotr and Belleville, Julie and Romero Cortés, Luis and Ruscitti, Kaleb and Fischer, Bennet and Azaña, José and Morandotti, Roberto},
title = {Inverse Design of Photonic Systems},
journal = {Laser \& Photonics Reviews},
volume = {18},
number = {5},
pages = {2300500},
doi = {https://doi.org/10.1002/lpor.202300500},
url = {https://onlinelibrary.wiley.com/doi/abs/10.1002/lpor.202300500},
year = {2024}
}

@article{PathIntegral,
    author = {Lackman-Mincoff, Jeremy and Jain, Moksh and Malkin, Nikolay and Bengio, Yoshua and Simine, Lena},
    title = {Path-filtering in path-integral simulations of open quantum systems using {GF}low{N}ets},
    journal = {The Journal of Chemical Physics},
    volume = {161},
    number = {14},
    pages = {144106},
    year = {2024},
    month = {10},
    issn = {0021-9606},
    doi = {10.1063/5.0226408},
    url = {https://doi.org/10.1063/5.0226408},
}

@article{
Mario_II,
author = {Xuemei Gu  and Manuel Erhard  and Anton Zeilinger  and Mario Krenn },
title = {Quantum experiments and graphs II: Quantum interference, computation, and state generation},
journal = {Proceedings of the National Academy of Sciences},
volume = {116},
number = {10},
pages = {4147-4155},
year = {2019},
doi = {10.1073/pnas.1815884116},
URL = {https://www.pnas.org/doi/abs/10.1073/pnas.1815884116},
}

@article{nonLocal-agents,
  title = {Automated discovery of nonlocal photonic gates},
  author = {Arlt, S\"oren and Krenn, Mario and Gu, Xuemei},
  journal = {Phys. Rev. Res.},
  volume = {8},
  issue = {2},
  pages = {L022031},
  numpages = {10},
  year = {2026},
  month = {May},
  publisher = {American Physical Society},
  doi = {10.1103/f415-kgwr},
  url = {https://link.aps.org/doi/10.1103/f415-kgwr}
}

@misc{GraphTransformerPyTorch,
      title={Masked Label Prediction: Unified Message Passing Model for Semi-Supervised Classification}, 
      author={Yunsheng Shi and Zhengjie Huang and Shikun Feng and Hui Zhong and Wenjin Wang and Yu Sun},
      year={2021},
      eprint={2009.03509},
      archivePrefix={arXiv},
      primaryClass={cs.LG},
      url={https://arxiv.org/abs/2009.03509}, 
}

@misc{GeneralizationTransformerOnGraphs,
      title={A Generalization of Transformer Networks to Graphs}, 
      author={Vijay Prakash Dwivedi and Xavier Bresson},
      year={2021},
      eprint={2012.09699},
      archivePrefix={arXiv},
      primaryClass={cs.LG},
      url={https://arxiv.org/abs/2012.09699}, 
}

@Article{DesignPhysExpAINature2026,
author={Klimesch, Jonathan
and Arlt, S{\"o}ren
and Ruiz-Gonzalez, Carlos
and Rodr{\'i}guez, Carla
and Gu, Xuemei
and Haslinger, Philipp
and Vischia, Pietro
and Haack, Christian
and Drori, Yehonathan
and Adhikari, Rana
and Arndt, Markus
and Kagan, Michael
and Heinrich, Lukas
and Krenn, Mario},
title={Designing physics experiments with artificial intelligence},
journal={Nature},
year={2026},
month={Sep},
day={01},
volume={657},
number={8130},
pages={47-58},
issn={1476-4687},
doi={10.1038/s41586-026-10898-6},
url={https://doi.org/10.1038/s41586-026-10898-6}
}

@misc{ArltAIMandel,
  author = {Arlt, S{\"o}ren and Gu, Xuemei and Krenn, Mario},
  title = {Towards autonomous quantum physics research using {LLM} agents with access to intelligent tools},
  year = {2025},
  eprint = {2511.11752},
  archivePrefix = {arXiv},
  primaryClass = {cs.AI},
  doi = {10.48550/arXiv.2511.11752},
  url = {https://arxiv.org/abs/2511.11752}
}

@article{ArltMetaDesign,
  author = {Arlt, S{\"o}ren and Duan, Haonan and Li, Felix and Xie, Sang Michael and Wu, Yuhuai and Krenn, Mario},
  title = {Meta-designing quantum experiments with language models},
  journal = {Nature Machine Intelligence},
  year = {2026},
  month = {Feb},
  volume = {8},
  number = {2},
  pages = {148--157},
  doi = {10.1038/s42256-025-01153-0},
  url = {https://doi.org/10.1038/s42256-025-01153-0}
}

@article{Raussendorf2003Cluster,
  author = {Raussendorf, Robert and Browne, Daniel E. and Briegel, Hans J.},
  title = {Measurement-based quantum computation on cluster states},
  journal = {Physical Review A},
  volume = {68},
  number = {2},
  pages = {022312},
  year = {2003},
  doi = {10.1103/PhysRevA.68.022312}
}

@article{Walther2005OneWay,
  author = {Walther, Philip and Resch, Kevin J. and Rudolph, Terry and Schenck, E. and Weinfurter, Harald and Vedral, Vlatko and Aspelmeyer, Markus and Zeilinger, Anton},
  title = {Experimental one-way quantum computing},
  journal = {Nature},
  volume = {434},
  pages = {169--176},
  year = {2005},
  doi = {10.1038/nature03347}
}

@article{Bartolucci2023Fusion,
  author = {Bartolucci, Sara and Birchall, Patrick and Bomb{\'i}n, Hector and Cable, Hugo and Dawson, Chris and Gimeno-Segovia, Mercedes and Johnston, Eric and Kieling, Konrad and Nickerson, Naomi and Pant, Mihir and Pastawski, Fernando and Rudolph, Terry and Sparrow, Chris},
  title = {Fusion-based quantum computation},
  journal = {Nature Communications},
  volume = {14},
  pages = {912},
  year = {2023},
  doi = {10.1038/s41467-023-36493-1}
}

@article{Knill2001LinearOptics,
  author = {Knill, E. and Laflamme, R. and Milburn, G. J.},
  title = {A scheme for efficient quantum computation with linear optics},
  journal = {Nature},
  volume = {409},
  pages = {46--52},
  year = {2001},
  doi = {10.1038/35051009}
}

@article{Liu2024Nonlocal,
  author = {Liu, Xiao and Hu, Xiao-Min and Zhu, Tian-Xiang and Zhang, Chao and Xiao, Yi-Xin and Miao, Jia-Le and Ou, Zhong-Wen and Li, Pei-Yun and Liu, Bi-Heng and Zhou, Zong-Quan and Li, Chuan-Feng and Guo, Guang-Can},
  title = {Nonlocal photonic quantum gates over 7.0 km},
  journal = {Nature Communications},
  volume = {15},
  pages = {8529},
  year = {2024},
  doi = {10.1038/s41467-024-52912-3}
}

@Article{LinearClusterEquiv,
author={Prevedel, Robert
and Walther, Philip
and Tiefenbacher, Felix
and B{\"o}hi, Pascal
and Kaltenbaek, Rainer
and Jennewein, Thomas
and Zeilinger, Anton},
title={High-speed linear optics quantum computing using active feed-forward},
journal={Nature},
year={2007},
month={Jan},
day={01},
volume={445},
number={7123},
pages={65-69},
issn={1476-4687},
doi={10.1038/nature05346},
url={https://doi.org/10.1038/nature05346}
}

@ARTICLE{Walther2005-nd,
  title     = "Experimental one-way quantum computing",
  author    = "Walther, P and Resch, K J and Rudolph, T and Schenck, E and
               Weinfurter, H and Vedral, V and Aspelmeyer, M and Zeilinger, A",
  journal   = "Nature",
  publisher = "Springer Science and Business Media LLC",
  volume    =  434,
  number    =  7030,
  pages     = "169--176",
  month     =  mar,
  year      =  2005,
}

@article{PhysRevLett.125.180502,
  title = {Entanglement-Assisted Absorption Spectroscopy},
  author = {Shi, Haowei and Zhang, Zheshen and Pirandola, Stefano and Zhuang, Quntao},
  journal = {Phys. Rev. Lett.},
  volume = {125},
  issue = {18},
  pages = {180502},
  numpages = {6},
  year = {2020},
  month = {Oct},
  publisher = {American Physical Society},
  doi = {10.1103/PhysRevLett.125.180502},
  url = {https://link.aps.org/doi/10.1103/PhysRevLett.125.180502}
}

@article{PhysRevLett.84.4729,
  title = {Quantum Cryptography with Entangled Photons},
  author = {Jennewein, Thomas and Simon, Christoph and Weihs, Gregor and Weinfurter, Harald and Zeilinger, Anton},
  journal = {Phys. Rev. Lett.},
  volume = {84},
  issue = {20},
  pages = {4729--4732},
  numpages = {0},
  year = {2000},
  month = {May},
  publisher = {American Physical Society},
  doi = {10.1103/PhysRevLett.84.4729},
  url = {https://link.aps.org/doi/10.1103/PhysRevLett.84.4729}
}
\end{document}